\documentclass[amsmath,amsfonts,superscriptaddress,twocolumn,prx,longbibliography]{revtex4-1}
\usepackage{graphicx}
\usepackage{epsfig}
\usepackage{bbm}
\usepackage{bm}
\usepackage{dcolumn}
\usepackage{amsmath}
\usepackage{amssymb}
\usepackage{color}
\usepackage{comment}
\usepackage[varg]{txfonts}

\newcommand{\beg}{\begin{equation}}
\newcommand{\en}{\end{equation}}
\newcommand{\bp}{\mathbf p}
\newcommand{\bq}{\mathbf q}

\newcommand{\br}{\mathbf r}

\newcommand{\bn}{\mathbf n}
\newcommand{\bx}{\mathbf x}

\newcommand \bel  {\begin{align}}
\newcommand \enl  {\end{align}}

\newcommand{\veps}{\varepsilon}
\newcommand{\eps}{\epsilon}

\begin{document}

\title{Spin Hall conductivity of interacting two-dimensional electron systems}

\author{Maxim Dzero}
\affiliation{Department of Physics, Kent State University, Kent, Ohio 44242, USA}

\author{Alex Levchenko}
\affiliation{Department of Physics, University of Wisconsin-Madison, Madison, Wisconsin 53706, USA}

\begin{abstract}
We consider a two-dimensional electron system subjected to a short-ranged nonmagnetic disorder potential, Coulomb interactions, and Rashba spin-orbit coupling. The path-integral approach incorporated within the Keldysh formalism is used to derive the kinetic equation for the semiclassical Green's function and applied to compute the spin current within the linear response theory. We discuss the frequency dependence of the spin Hall conductivity and further elucidate the role of electron interactions at finite temperatures for both the ballistic and diffusive regimes of transport. We argue that interaction corrections to the spin Hall effect stem from the quantum interference processes whose magnitude is estimated in terms of parameters of the considered model.  \\

\textit{Spin-Coherent Phenomena in Semiconductors: Special Issue in Honor of E. I. Rashba}. 
\end{abstract}

\date{November 19, 2022}

\maketitle

\section{Introduction}  

At the present day the physics phenomena originating from the Rashba spin-orbit coupling \cite{Rashba1960,Bychkov1984} dominate the fields of spintronics \cite{Dyakonov2017} and topological systems \cite{Shen2017}. One fundamentally important example of the quantum spin transport that intertwines with topological properties of the band structure is the spin Hall effect (SHE). It represents a collection of transport phenomena
whereby charge currents propagating in nonmagnetic materials are converted to transverse spin currents and vice versa. The resulting spin current is even under the application of the time-reversal operator, so that in this regard this effect is similar to the appearance of dissipationless current in $s$-wave superconductors. 
Studies of the extrinsic mechanisms of the SHE have rich history dating back to the original works \cite{Dyakonov-Perel,Hirsch}; see also review \cite{Sinova-RMP} and references therein. In general, this effect has conceptual overlap with the anomalous Hall effect (AHE) \cite{Nagaosa}, where one distinguishes extrinsic side-jump and skew-scattering mechanisms of the transverse responses, as well as an intrinsic source generated by the Berry curvature in materials with topologically nontrivial band structure \cite{Niu}.  

\begin{figure}[t!]
\centering
\includegraphics[width=0.9\linewidth]{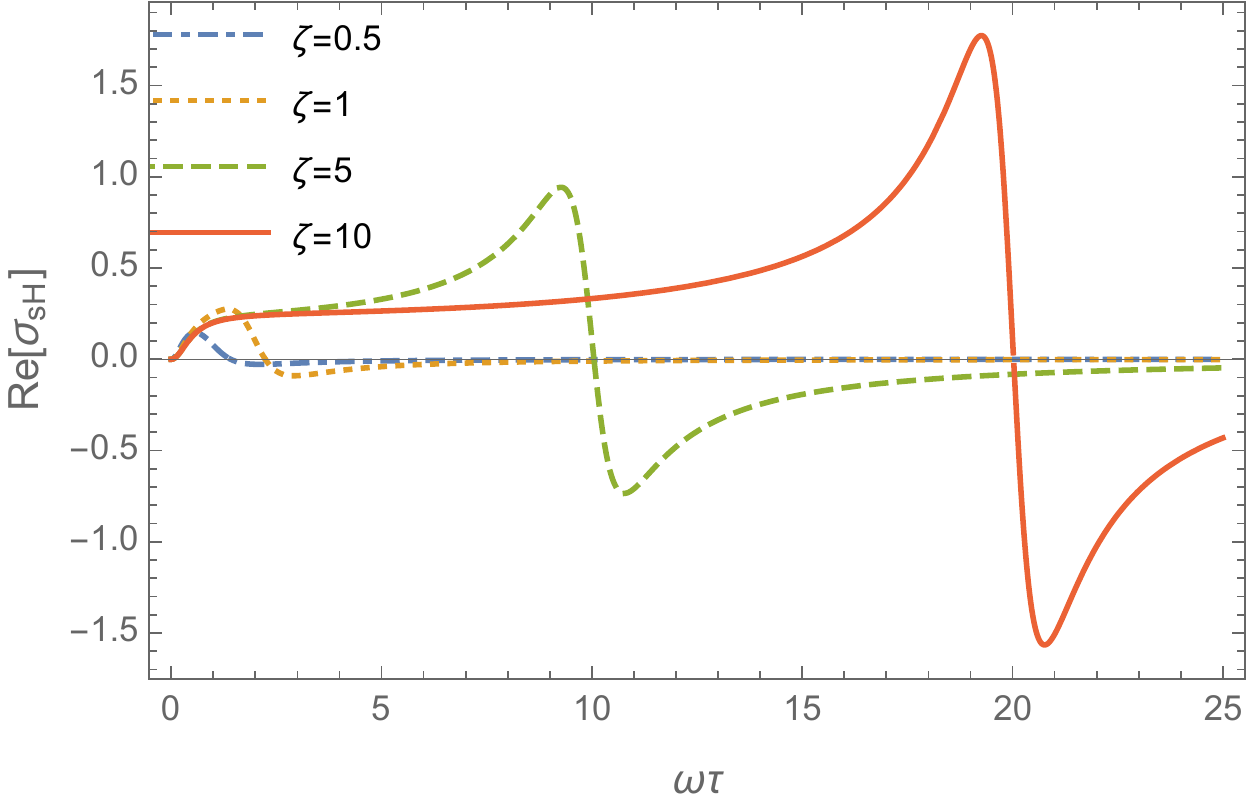}
\includegraphics[width=0.9\linewidth]{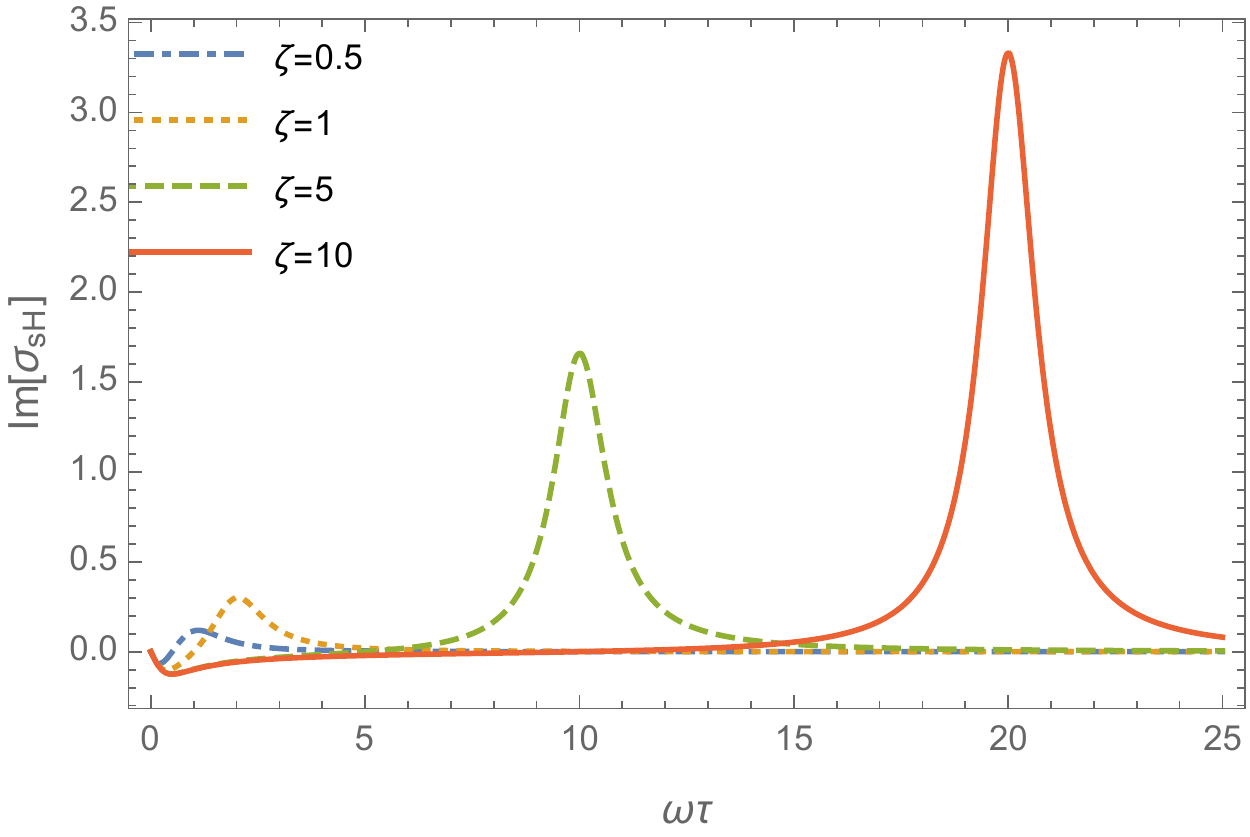}
\caption{Frequency dependence of the spin Hall conductivity in the Rashba model of spin-orbit coupling for 2DES subjected to the short-ranged quenched disorder potential characterized by the elastic scattering time $\tau$. Plots of the real part (a) and imaginary part (b) are made for the different strength of the disorder potential as characterized by the parameter $\zeta=\Delta\tau$, where $\Delta=\lambda_{\text{so}}p_F$ is the energy scale of spin-orbit-induced band splitting.}
\label{Fig-ReIm-sigma}
\end{figure}

In the SHE the transverse $z$-component spin current is induced by an electric in-plane field through the spin-orbit coupling, 
\begin{equation}\label{eq:jz}
j^z_i=\sigma_{\text{sH}}\epsilon_{ij}\mathcal{E}_j,
\end{equation}
and therefore one would naturally expect that the corresponding spin Hall conductivity should depend on the value of the spin-orbit coupling $\lambda_{\text{so}}$.
In the inspiring work by Sinova \textit{et al.} \cite{Sinova2004} it was proposed that the spin current may appear in $n$-type spin-orbit coupled semiconductors, furthermore, it was shown that in the bulk of a clean and homogeneous two-dimensional electronic system (2DES), the spin Hall conductivity is reactive and predicted to have a universal value $\sigma_{\textrm{sH}}=e/8\pi$ independent of the spin-orbit coupling (in units of $\hbar=1$). This specific result pertains to the Rashba model of the spin-orbit interaction. The universality is bounded by the condition on the carrier concentration that must exceed a threshold, $n>(m\lambda_{\text{so}})^2/\pi$, where $m$ is an effective mass, so that both spin-split bands are occupied. This is not a unique example, as universal SHE was predicted also to occur in the Luttinger model Hamiltonian \cite{Zhang2003,Zhang2004}. 

These works generated significant interest and ignited intense research on this peculiar transport effect \cite{Rashba2003,Culcer2004,Shen2004,Sinitsyn2004,Khodas2005}, scrutinizing particularly the robustness of the universality of spin Hall conductivity \cite{Rashba2004,Murakami2004,Schliemann2004,Inoue2004,MSH2004,Burkov2004,Raimondi2005,Olga2005,Chalaev2005,Bernevig2005,Onoda2005,Sheng2005,Chen2005,Nomura2005,Engel2005,Sugimoto2006,Raimondi2006,Khaetskii2006}. The resulting conclusions seem to differ between the model cases. In the Luttinger model an intrinsic spin Hall conductivity remains largely intact by scattering \cite{Murakami2004,Schliemann2004,Bernevig2005,Onoda2005,Chen2005}, whereas in the Rashba model inclusion of even the smallest amount of nonmagnetic disorder with the short-range impurity potential leads to the vanishing of the dc spin Hall conductivity \cite{Inoue2004,MSH2004,Raimondi2005,Olga2005,Chalaev2005,Khaetskii2006}. This is due to the fact that scattering introduces the dissipative contribution to the spin Hall conductivity $\sigma_{\textrm{sH}}^{D}=-e/8\pi$ that cancels out the reactive part $\sigma_{\textrm{sH}}^R=e/8\pi$. 
This result was verified independently by employing a variety of complementary techniques that include the diagrammatic Kubo formula approach \cite{Inoue2004,Raimondi2005,Olga2005,Chalaev2005}, Boltzmann equation \cite{Khaetskii2006}, semiclassical Eilenberger kinetic equation \cite{MSH2004,Raimondi2006}, and direct numerical finite-size analysis \cite{Sheng2005,Nomura2005}. With diagrams the cancellation can be traced to the delicate interplay of self-energy effect and current vertex corrections. However, despite the general consensus questions still remained whether this is just specific properties of the linear Rashba model and perhaps an artifact of the model assumptions on the properties of the disorder potential. 

Further detailed analyses suggest the following. (1) In the delta-function disorder model the weak-localization corrections still yield vanishing dc spin Hall conductivity \cite{Chalaev2005}; however, going beyond Born approximation and incorporating certain diffractive scattering processes with crossed impurity lines may render finite contributions \cite{Ferreira2016}. (2) This cancellation is also not complete in an extended model with the finite range impurity potential \cite{Sugimoto2006} and requires a refined definition of the spin current operator \cite{Shi2006}. (3) The dc spin Hall conductivity also remains finite in the disordered artificially engineered contacts and/or heterostructures \cite{MSH2004,Abanin2009,Funato2020}. (4) Concerning the impact of electron interaction, general arguments were put forward to suggest that vanishing of the dc spin current is an exact property of the Rashba 2DES with any nonmagnetic interaction \cite{Olga2005,Chalaev2005}. The argument is based on the observation that the Heisenberg equation of motion for the spin $\hat{\mathbf{s}}=\hat{\bm{\sigma}}/2$ of the electron, $\dot{\hat{\mathbf{s}}}=i[\hat{H},\hat{\mathbf{s}}]$, gives a simple relation between the spin precession and the spin current 
\begin{equation}\label{eq:dot-s}
\dot{\hat{s}}_i(t)=-2m\lambda_{\text{so}}\hat{j}^z_i(t),\quad i=x,y.
\end{equation}
The presence of disorder provides a relaxation mechanism so that the system is expected to reach a steady state in a long-time evolution where the expectation value of magnetization $\langle\hat{\mathbf{s}}\rangle$ reaches some stationary value. Therefore, the rate of spin polarization change, $\langle\dot{\hat{\mathbf{s}}}\rangle$, must vanish,   
and as an immediate consequence of Eqs. \eqref{eq:jz} and \eqref{eq:dot-s} $\sigma_{\text{sH}}=0$ for the static electric field. In contrast,  in clean systems it was found that the long-range or screened electron-electron interactions will modify the universal value of $\sigma_{\textrm{sH}}^R$ \cite{Olga2005}. (5) Finally, the ac spin Hall conductivity remains finite due to its relation with the Pauli susceptibility \cite{Olga2005,Khodas2005}.

In order to reconcile the initial prediction of Ref. \cite{Sinova2004} with the rest of the studies that followed on the Rashba model with short-ranged disorder it is most instructive to study spin Hall conductivity at finite frequency. It becomes then clear that orders of taking the clean limit first or sending frequency to zero first do not commute. Indeed, keeping the impurity scattering time $\tau$ finite and sending the external frequency to zero gives $\sigma_{\text{sH}}=(e/4\pi i)\omega\tau$, thus indeed resulting in vanishing dc spin Hall conductivity. Instead, keeping frequency finite but taking a ballistic limit with $\tau\to\infty$ gives universal $\sigma_{\text{sH}}=e/8\pi$, which happens to be frequency independent, but this is not a dc expression. Strictly speaking the saturated value of $\sigma_{\text{sH}}$ occurs at the intermediate range of frequencies, $\tau^{-1}_s<\omega<\tau^{-1}$, between the rates of spin relaxation and elastic scattering where it is given by $\sigma_{\text{sH}}=(e/4\pi)(\tau/\tau_s)$. The ratio $\tau/\tau_s$ is strongly suppressed in the diffusive limit by the Dyakonov-Perel mechanism of spin relaxation \cite{DP1971}, at the same time it saturates to $1/2$ in the ballistic regime. Finally, at highest frequencies $\omega\tau\gg1$ ac spin Hall conductivity decays algebraically as $\sigma_{\text{sH}}=(e/2\pi)(1/\omega\tau)^{2}$. This behavior is further illustrated in Fig. \ref{Fig-ReIm-sigma} for both real and imaginary parts of $\sigma_{\text{sH}}(\omega)$ and an analytical formula is derived later in the paper. 

Despite all these research efforts and to the best of our knowledge the general problem of an interplay between the electron-electron interactions and disorder on spin current has not been fully addressed yet. In this work, we attempt in part to fill this gap. Specifically, we take the general theoretical framework developed earlier in Refs. \cite{MSH2004,Raimondi2006,LiLi2008} one step further and study the problem of the effects of Coulomb interactions on spin Hall conductivity in the presence of nonmagnetic disorder scattering. We find the quantum interference correction to the spin Hall conductivity at finite temperatures.  
The physical origin of the interaction correction is analogous to the Al'tshuler-Aronov effect in electrical conductivity of disordered 2DES and stems from the interference between the two semiclassical electronic paths: the first path corresponds to a scattering on impurity, while the second one corresponds to a scattering off the Friedel oscillations created by an impurity \cite{Rudin1997,ANZ2001}. This interaction correction appears in the $1/(p_Fl)$ order with $p_F$ the Fermi momentum and $l$ the mean-free path. It shows linear in temperature dependence, which is further independent of the strength of spin orbit in the ballistic regime.

The rest of the paper is organized as follows. In Sec. \ref{sec:Model} we introduce model of interacting 2DES with spin-orbit coupling subjected to the disorder potential and present main ingredients of the formalism. In Sec. \ref{sec:KinEq} we derive the kinetic equation for the quasiclassical Green's function. An equation for the   
density matrix is obtained in Sec. \ref{sec:DensityMatrix}. The spin current response to the alternating field is analyzed in Sec. \ref{sec:SpinCurrent} and interaction corrections to the spin Hall conductivity are estimated. Technical calculations of the collision integrals are delegated to several Appendixes.  

%#############################################################################################################################
%#############################################################################################################################
%#############################################################################################################################
%#############################################################################################################################
%#############################################################################################################################

\section{Model}\label{sec:Model}

In the following we provide the technical ingredients of our theory. The detailed introduction to the main theoretical framework can be found in the original paper by Zala \textit{et al.} \cite{ANZ2001}. The generalization of their results to a disordered system of the two-dimensional interacting Fermi gas with Rashba spin-orbit coupling has been given by Li and Li \cite{LiLi2008}. Our technical discussion here can be viewed as an extension of the results of Ref. \cite{LiLi2008} by retaining in the kinetic equation the terms linear in powers of $\lambda_{\text{so}}/v_F$. Lastly, we note that in what follows we adopt the energy units $\hbar=c=k_B=1$. 

We start with the partition function in the Keldysh formulation of nonequilibrium systems \cite{AKAL2009,Kamenev}
\beg\label{Z}
{\cal Z}=\int D[\overline{\psi}\psi]\exp\left(-iS[\overline{\psi},\psi]\right),
\en
where $\overline{\psi}_\alpha$, $\psi_\alpha$ are fermionic Grassman fields, $\alpha$ is a spin label, and $S[\overline{\psi},\psi]$ is an action defined as
\beg\label{S}
\begin{split}
S[\overline{\psi},\psi]&=\int_Cdt\int d^2\br\sum\limits_{\alpha\beta}\overline{\psi}_\alpha(\br,t)
\hat{\cal L}_{\alpha\beta}(\br,t){\psi}_\beta(\br,t)\\&+\frac{1}{\cal A}\int\limits_{-\infty}^{\infty} dt\sum\limits_{\bq\not=0}\frac{\pi e^2}{|\bq|}\hat{\varrho}(\bq,t)\hat{\varrho}(-\bq,t).
\end{split}
\en
In this expression the time integration is performed over the Keldysh contour, $\hat{\varrho}(\bq,t)$ is the Fourier transform of the particle density operator $\hat{\varrho}(\br,t)=\sum_\alpha\overline{\psi}_\alpha(\br,t){\psi}_\alpha(\br,t)$, ${\cal A}=L^2$, and $L$ is a characteristic linear size of the sample. $\hat{\cal L}$ is an operator defined in the two-by-two matrix in spin space
\beg\label{hab}
\hat{\cal L}(\br,t)=\left[-i\frac{\partial}{\partial t}-\frac{{\bm{\nabla}}^2}{2m}-E_F+U(\br)\right]\hat{\mathbbm{1}}_{2\times 2}+\hat{\eta}_{\bp},
\en
where $E_F$ is the Fermi energy, $\hat{\eta}_\bp=\lambda_{\text{so}}[{\bm{e}}_z\times\hat{{\bm{\sigma}}}]\cdot(-i\bm{\nabla})$, $\lambda_{\text{so}}$ is the spin-orbit coupling, and $U(\br)$ is the random disorder potential with the correlator 
\beg\label{Disorder}
\langle U(\br)U(\br')\rangle_{\textrm{dis}}=\frac{1}{2\pi\nu_F\tau}\delta(\br-\br').
\en
Averaging $\langle...\rangle_{\textrm{dis}}$ is performed over disorder realizations, and $\nu_F$ is the density of states at the Fermi level. 

Since the physical observables are generally expressed in terms of the fermionic correlators we consider the single-particle Green's function
\beg\label{Gxxp}
\hat{G}_{\alpha\beta}(x,x')=\frac{-i}{\cal Z}\int D[\overline{\psi}\psi]\hat{T}_C\psi_{\alpha}(x)\overline{\psi}_\beta(x')e^{-iS[\overline{\psi},\psi]}.
\en
and $T_C$ refers to the time ordering operator on a Keldysh contour.

%$$$$$$$$$$$$$$$$$$$$$$$$$$$$$$$$$$$$$$$$
\subsection{Hubbard-Stratonovich transformation}
%$$$$$$$$$$$$$$$$$$$$$$$$$$$$$$$$$$$$$$$$

Our goal is to obtain an equation for the Green's function. This is done in two steps. The first step consists in performing the Hubbard-Stratonovich transformation:
\beg\label{HSTransform}
\begin{split}
&\exp\left(-\frac{i}{\cal A}\int\limits_Cdt\sum\limits_{\bq\not=0}\frac{\pi e^2}{|\bq|}\hat{{\varrho}}(\bq,t)\hat{{\varrho}}(-\bq,t)\right)\\&=
\int D[\phi]\exp\left[i\int_Cdt\sum\limits_{\bq}\left(\frac{|\bq|}{4\pi}\right)\phi(\bq,t)\phi(-\bq,t)\right]\times \\ &
\exp\left\{\frac{ie}{2\sqrt{\cal A}}\int_Cdt\sum\limits_{\bq}[\hat{{\varrho}}(\bq,t)\phi(-\bq,t) 
+\hat{{\varrho}}(-\bq,t)\phi(\bq,t)]\right\}.
\end{split}
\en
In the second step, we consider separately the fermionic and bosonic fields which reside on two sides of the Keldysh contour: $\overline{\psi}_{i\alpha}$,
$\psi_{i\alpha}$ and $\phi_i$ with $i=+$ for the top part of the contour and $i=-$ for the bottom part, so that we can treat the fermionic fields as doublets in the Keldysh space:
\beg\label{Doublets}
\Psi_\alpha=\left(\begin{matrix} \psi_{\alpha,{+}} \\ \psi_{\alpha,{-}} \end{matrix}\right), \quad \Phi=\left(\begin{matrix} \phi_{+} \\ \phi_{-} \end{matrix}\right).
\en
The original action in Eq. \eqref{S} can now be written as a sum of the term quadratic in the fermionic fields and a term which is purely bosonic: $S[\overline{\psi},\psi;\phi]=S_{\textrm{f-b}}[\overline{\psi},\psi;\phi]+S_{\textrm b}[\phi]$ with
\beg\label{Actions}
\begin{split}
S_{\textrm{f-b}}&=\int_x\overline{\Psi}_{\alpha}(x)\left[\hat{\cal L}_{\alpha\beta}(x)\hat{\tau}_3-e\hat{\gamma}_i\phi_i(x)\delta_{\alpha\beta}\right]\Psi_\beta(x), \\
S_{\textrm{b}}&=-\frac{e^2}{2}\int_x\int_{x'}\Phi^T(x)V^{-1}(x-x')\hat{\tau}_3\Phi(x').
\end{split}
\en 
Here the summation over repeated indices is assumed. Further, we use the shorthand notations: $x=(\br,t)$, $\int_x=\int_{-\infty}^\infty dt\int d^2\br$ and matrices
\beg\label{AuxMatrix}
\hat{\tau}_3=\left(\begin{matrix} 1 & 0 \\ 0 & -1 \end{matrix}\right), \quad 
\hat{\gamma}_{+}=\left(\begin{matrix} 1 & 0 \\ 0 & 0 \end{matrix}\right), \quad 
\hat{\gamma}_{-}=\left(\begin{matrix} 0 & 0 \\ 0 & -1 \end{matrix}\right),
\en
which all operate in the Keldysh space. The function $V^{-1}(x-x')$ is defined according to $V^{-1}(x-x')=V^{-1}(\br-\br')\delta(t-t')$ and $V^{-1}(\br)$ can be found by solving the equation
\beg\label{InvV}
e^2\int V^{-1}(\br-\br_1)\frac{d^2\br_1}{|\br_1-\br'|}=\delta(\br-\br').
\en

The bosonic fields are described by the correlators
\beg\label{Dab}
{\cal D}_{ab}(x,x')=i\langle\phi_a(x)\phi_b(x')\rangle.
\en
Let us single out the second term in $S_{\textrm{f-b}}[\overline{\psi},\psi;\phi]$ and denote it by $S_{\textrm{int}}[\overline{\psi},\psi;\phi]$. Introducing the partition function corresponding to a fixed configuration of the bosonic fields
\beg\label{Zphi}
{Z}[\phi]=\int D[\overline{\psi},\psi]{T}_C\exp(-iS_{\textrm{int}}[\overline{\psi},\psi;\phi]),
\en
we consider the corresponding fermionic Green's function for this fixed configuration 
\beg\label{Gabphi}
\begin{split}
\hat{G}_{\alpha\beta}(x,x'\vert\phi)=\frac{-i}{Z[\phi]}&\int D[\overline{\psi}\psi]\hat{T}_C\Psi_{\alpha}(x)\overline{\Psi}_\beta(x')\\ &\times 
e^{-iS_{\textrm{f-b}}[\overline{\psi},\psi;\phi]}.
\end{split}
\en
Here $\hat{G}(x,x')$ is a $4\times 4$ matrix in Keldysh and spin spaces, $G_{\alpha\beta}^{(ij)}(x_i,x_j')$ $(i,j=\pm)$. It then follows
\beg\label{Gab}
\hat{G}_{\alpha\beta}(x,x')=\frac{1}{\cal Z}\int D[\phi] \hat{G}_{\alpha\beta}(x,x'\vert\phi)e^{-iS_{\textrm{B}}[\phi]}, 
\en
where $S_{\textrm{B}}[\phi]=S_{\textrm{b}}[\phi]+i\ln Z[\phi]$. Averaging both parts of this expression over disorder yields
\beg\label{disavGab}
\langle\hat{G}_{\alpha\beta}(x,x')\rangle_{\textrm{dis}}=\frac{1}{\cal Z}\int D[\phi]\langle\hat{G}_{\alpha\beta}(x,x'\vert\phi)\rangle_{\textrm{dis}}
e^{-i\langle S_{\textrm{B}}[\phi]\rangle_{\textrm{dis}}}.
\en
This expression is approximate since it ignores mesoscopic fluctuations between polarizability of the medium due to the action of the bosonic fields  and conduction electrons: the effect of these fluctuations is of the order of $\propto(E_F\tau)^{-2}$ \cite{ANZ2001}.

%$$$$$$$$$$$$$$$$$$$$$$$
\subsection{Basic equations}
%$$$$$$$$$$$$$$$$$$$$$$$

We proceed to introduce quantities which will be central to our subsequent discussion. Our presentation below follows closely 
the discussion in Ref. \cite{ANZ2001}. We treat the bosonic action $S_{\textrm{B}}[\phi]$ in the saddle-point approximation:
\beg\label{FBphi}
\begin{split}
F[\phi]=i\ln \langle e^{-iS_{\textrm{B}}[\phi]}\rangle=F[0]+\frac{1}{2}\phi^T\hat{\Pi}\phi+O(\phi^3),
\end{split}
\en
where $\langle...\rangle$ formally denotes the averaging over bosonic degrees of freedom around the saddle-point value and $\hat{\Pi}$ is the electronic polarization operator
\beg\label{Pi}
\Pi_{\alpha\beta}(x,x')=\left(\frac{\delta^2F[\phi]}{\delta\phi_\alpha(x)\delta\phi_\beta(x')}\right)_{\phi=0}.
\en
In order to express the polarization operator in terms of the fermionic propagator (\ref{Gabphi}) it would be 
convenient to perform the Keldysh rotation for the fermionic fields (we suppress the spin label for brevity):
\beg\label{KeldyshRotate}
\begin{split}
&\psi_{1}=\frac{1}{\sqrt{2}}(\psi_{+}+\psi_{-}), \quad {\psi}_{2}=\frac{1}{\sqrt{2}}(\psi_{+}-\psi_{-}), \\
&\overline{\psi}_{1}=\frac{1}{\sqrt{2}}(\overline{\psi}_{+}-\overline{\psi}_{-}), \quad 
\overline{\psi}_{2}=\frac{1}{\sqrt{2}}(\overline{\psi}_{+}+\overline{\psi}_{-}).
\end{split}
\en
The corresponding relations for the bosonic fields are $\phi_{1(2)}=(\phi_{+}\pm\phi_{-})/2$. As a result of this rotation, the Green's function in Eq. (\ref{Gabphi}) has the form
\beg\label{RotatedG}
\hat{G}_{\alpha\beta}(x,x'\vert\phi)=\left(\begin{matrix} 
G_{\alpha\beta}^R(x,x'\vert\phi) \quad G_{\alpha\beta}^K(x,x'\vert\phi) \\
G_{\alpha\beta}^Z(x,x'\vert\phi) \quad G_{\alpha\beta}^A(x,x'\vert\phi)
\end{matrix}
\right).
\en
In accordance with the standard convention $G^{R/A/K}$ denotes retarded/advanced/Keldysh Green's function.  
It is important to emphasize that $G_{\alpha\beta}^Z(x,x'\vert\phi)$ by itself is not zero, but its average over the bosonic fields must be zero by causality, $\langle G_{\alpha\beta}^Z(x,x'\vert\phi)\rangle_\phi=0$.

The bosonic propagators (\ref{Dab}) in the rotated basis are defined as follows:
\beg\label{BosonicProps}
\begin{split}
&\langle\phi_1(x)\phi_1(x')\rangle=\frac{i}{2}{\cal D}^K(x,x'), \\
&\langle\phi_1(x)\phi_2(x')\rangle=\frac{i}{2}{\cal D}^R(x,x'), \\
&\langle\phi_2(x)\phi_1(x')\rangle=\frac{i}{2}{\cal D}^A(x,x'), ~\langle\phi_2(x)\phi_2(x')\rangle=0.
\end{split}
\en
From Eqs. (\ref{FBphi}) we find that if we were to ignore $S_{\textrm{f-b}}$, then ${\cal D}_0^A(x,x')={\cal D}_0^R(x,x')=-V(\br-\br')\delta(t-t')$
and ${\cal D}_0^K(x,x')=0$. With the help of the saddle-point approximation (\ref{FBphi}), the Dyson equation for the bosonic propagators can be compactly written as
\beg\label{BosonicDyson}
\begin{split}
&\hat{\cal D}(x,x')=\hat{\cal D}_0(x,x')\\&+\int dx_3\int dx_4 \hat{\cal D}_0(x,x_3)\hat{\Pi}(x_3,x_4)\hat{\cal D}(x_4,x'), \\
&\hat{\cal D}=\left(\begin{matrix} 
 {\cal D}^R & {\cal D}^K \\
       0         & {\cal D}^A 
\end{matrix}
\right), \quad \hat{\Pi}=\left(\begin{matrix} 
 {\Pi}^R & {\Pi}^K \\
       0         & {\Pi}^A 
\end{matrix}
\right).
\end{split}
\en
Furthermore, from expanding $Z[\phi]$ around the saddle point, we obtain the following relations between the polarization operators and 
the fermionic correlators:
\beg\label{PiGRelations}
\begin{split}
\Pi^R(x_1,x_2)&=\left(\frac{i}{2}\right)\frac{\delta}{\delta\phi_1(x_2)}\left\{\textrm{Tr}_\sigma\left[\hat{G}^K(x,x|\phi)\right]\right\}, \\
\Pi^K(x_1,x_2)&=\left(\frac{i}{2}\right)\frac{\delta}{\delta\phi_2(x_2)}\left\{\textrm{Tr}_\sigma\left[\hat{G}^K(x,x|\phi)\right]\right\}\\&+
\left(\frac{i}{2}\right)\frac{\delta}{\delta\phi_2(x_2)}\left\{\textrm{Tr}_\sigma\left[\hat{G}^Z(x,x|\phi)\right]\right\},
\end{split}
\en
and $\Pi^A(x,x')=\Pi^R(x',x)$.

It is well known that fermionic Green's function satisfies the pair of Dyson equations \cite{Kamenev}:
\beg\label{Dyson4Ferms}
\left(\hat{G}_0^{-1}-\hat{\Sigma}\right)\circ\hat{G}={\mathbbm{1}}, \quad \hat{G}\circ\left(\hat{G}_0^{-1}-\hat{\Sigma}\right)={\mathbbm{1}}.
\en
In these equations the matrix product implies the convolution in time and space as well as the matrix product in spin and Keldysh spaces, and the operator $\hat{G}_0$ is given by
\beg\label{G0}
\begin{split}
\hat{G}_0^{-1}&=\left[i\frac{\partial}{\partial t}+\frac{{\bm{\nabla}}^2}{2m}+E_F-U(\br)-\hat{\phi}(x)\right]\hat{\mathbbm{1}}_{2\times 2}\\&-
\lambda_{\text{so}}[{\bm{e}}_z\times\hat{{\bm{\sigma}}}]\cdot(-i{\bm{\nabla}}), 
\quad \hat{\phi}=\left(\begin{matrix} \phi_1 & \phi_2 \\ \phi_2 & \phi_1\end{matrix}\right).
\end{split}
\en
We can now perform the averaging over disorder in the leading in the $1/(E_F\tau)$ approximation. This procedure transforms the first equation (\ref{Dyson4Ferms}) to the following form:
\beg\label{Eq4Ga}
\begin{split}
&\hat{h}(1)\hat{G}(x_1,x_2\vert\phi)-\hat{\phi}(x_1)\hat{G}(x_1,x_2\vert\phi)\\&-
\lambda_{\text{so}}[\bm{e}_z\times\hat{\bm{\sigma}}]\cdot(-i\bm{\nabla}_1)\hat{G}(x_1,x_2\vert\phi)\\&=
\hat{\mathbbm{1}}\delta(x_1-x_2)+\int dx_3\hat{\Sigma}(x_1,x_3\vert\phi)\hat{G}(x_1,x_2\vert\phi),
\end{split}
\en
where $\hat{h}(1)=i\partial/\partial t_1+\bm{\nabla}_1^2/(2m)+E_F$ 
and the self-energy part, on the account of 
Eq. (\ref{Disorder})), is given by
\beg\label{SelfEnergy}
\hat{\Sigma}(x_1,x_2\vert\phi)=\frac{\delta(\br_1-\br_2)}{2\pi\nu_F\tau}\hat{G}(x_1,x_2\vert\phi).
\en
Consequently, the second equation (\ref{Dyson4Ferms}) averaged over disorder reads:
\beg\label{Eq4Gb}
\begin{split}
&\hat{h}^*(2)\hat{G}(x_1,x_2\vert\phi)-
\hat{G}(x_1,x_2\vert\phi)\hat{\phi}(x_2)\\&-
\lambda_{\text{so}}\left(i\bm{\nabla}_2\hat{G}(x_1,x_2\vert\phi)\right)[\bm{e}_z\times\hat{\bm{\sigma}}]\\&=
\hat{\mathbbm{1}}\delta(x_1-x_2)+\int dx_3\hat{G}(x_1,x_3\vert\phi)
\hat{\Sigma}(x_3,x_2\vert\phi).
\end{split}
\en
Formally, Eqs. (\ref{Eq4Ga}( and (\ref{Eq4Gb}) together with the relations (\ref{PiGRelations}) and (\ref{BosonicDyson}) form a closed set. Notice that Eqs. (\ref{Eq4Ga}) and (\ref{Eq4Gb}) include the bosonic field explicitly and not the bosonic correlators. However, it would be desirable to transform (\ref{Eq4Ga}) and (\ref{Eq4Gb}) into the form which would allow us to express $\hat{G}$ directly in terms of $\hat{\cal D}$. This goal can be accomplished in several steps. The first step consists in treating the equations of motion (\ref{Eq4Ga}) and (\ref{Eq4Gb}) semiclassicaly \cite{Semi1,Semi2,Semi3}. 

%$$$$$$$$$$$$$$$$$$$$$$$$$$$
\subsection{Eilenberger equation}
%$$$$$$$$$$$$$$$$$$$$$$$$$$$

At this point we introduce the quasiclassical Green's function. It is obtained from the function $\hat{G}(x_1,x_2\vert\phi)$, which has already been averaged over disorder, by averaging over the distance from the Fermi surface:
\beg\label{smallg}
\hat{g}(t_1,t_2;\bn,\br)=\frac{i}{\pi}\int\limits_{-\infty}^{+\infty}\hat{G}\left(t_1,t_2;\bp(\xi),\br\vert\phi\right)d\xi,
\en
where $\bp(\xi)\approx (\bp/p)\sqrt{p_F+\xi/v_F}$ and $\br=(\br_1+\br_2)/2$. Following the avenue of Refs. \cite{Semi2,Semi3,ANZ2001,LiLi2008}, we subtract Eq. (\ref{Eq4Gb}) from Eq. (\ref{Eq4Ga}) and keep only linear in spatial gradient terms assuming that $\hat{g}(t_1,t_2;\bn,\br)$ varies slowly with $\br$. The resulting equation is the Eilenberger equation for the quasiclassical function $\hat{g}(t_1,t_2;\bn,\br)$:
\beg\label{Eilen1}
\begin{split}
&\bm{\partial}_t\hat{g}+v_F(\bn\cdot\bm{\nabla})\hat{g}+\frac{\lambda_{\text{so}}}{2}\left\{{\hat{\bm{\eta}}},\bm{\nabla}\hat{g}\right\}
+i\Delta_p\left[\hat{\eta}_\bp,\hat{g}\right]\\&=
\frac{1}{2\tau}\left(\hat{g}\langle \hat{g}\rangle_\bn-\langle \hat{g}\rangle_\bn\hat{g}\right).
\end{split}
\en
The product of the functions on the right-hand side should be understood as convolution in time and product in spin and Keldysh spaces. 
On the left-hand side of this equation the curly brackets denote the anticommutation relation while the square brackets denote the commutator. Also,
we consider parameter $\Delta_p=\lambda_{\text{so}}p$, where $\langle...\rangle_\bn$ stands for the averaging over the directions of momentum, 
\beg\label{partt}
\bm{\partial}_t\hat{g}=\frac{\partial \hat{g}}{\partial t_1}+\frac{\partial \hat{g}}{\partial t_2}+i\hat{\phi}(\br,t_1)\hat{g}-i\hat{g}\hat{\phi}(\br,t_2)
\en
and $\hat{\bm{\eta}}=[\bm{e}_z\times\hat{\bm{\sigma}}]$. It is important to keep in mind, that the solutions of Eq. \eqref{Eilen1} are subjected to the constraints \cite{ANZ2001}
\begin{subequations}\label{Constraint}
\begin{align}
&\hat{g}(\bn,\br)\hat{g}(\bn,\br)\nonumber \\
&=\int\limits_{-\infty}^\infty dt_3\sum\limits_l[\hat{g}(t_1,t_3;\bn,\br)]_{il}[\hat{g}(t_3,t_2;\bn,\br)]_{lj}=\mathbbm{1}^K, \\
&\int\limits_{-\infty}^\infty dt\textrm{Tr}\left[\hat{g}(t,t;\bn,\br)\right]=0.
\end{align}
\end{subequations}
and $\hat{g}$ has the same matrix form in the Keldysh space as (\ref{RotatedG}).
We note that the right-hand side of Eq. \eqref{Eilen1} contains contributions in the parameter $\lambda_{\text{so}}/v_F$ that are important for the calculation of the spin current (see Sec. \ref{sec:DensityMatrix}). 
These terms can be neglected if one is only interested in studying the effects of interactions on dynamics of the spin relaxation \cite{LiLi2008}.

%#############################################################################################################################
%#############################################################################################################################
%#############################################################################################################################
%#############################################################################################################################
%#############################################################################################################################

%$$$$$$$$$$$$$$$$$$$$$$$$$$$$$$$$$$
\section{Kinetic equation}\label{sec:KinEq}
%$$$$$$$$$$$$$$$$$$$$$$$$$$$$$$$$$$

The physical observables are determined by the Keldysh component of the quasiclassical function $\hat{g}$ \cite{Kamenev}. Therefore, our next task will be to derive the kinetic equation for $\hat{g}^K$ averaged over the bosonic fields, $\langle\hat{g}^K\rangle_\phi$. The equation for $\hat{g}^K$ can be obtained from the corresponding matrix block of Eq. (\ref{Eilen1}):
\beg\label{Eq4gK}
\begin{split}
&\left[{\partial}_t+v_F(\bn\cdot \bm{\nabla})\right]\hat{g}^K+\frac{\lambda_{\text{so}}}{2}\left\{\hat{\bm{\eta}},\bm{\nabla}\hat{g}^K\right\}\\&+i\Delta_p\left[\hat{\eta}_\bp,\hat{g}^K\right]=
-i[\phi_1(\br,t_1)-\phi_1(\br,t_2)]\hat{g}^K\\&-i\phi_2(\br,t_1)\hat{g}^A+i\phi_2(\br,t_2)\hat{g}^R\\&
+\frac{1}{2\tau}\left(\hat{g}^R\langle \hat{g}^K\rangle_\bn+\hat{g}^K\langle \hat{g}^A\rangle_\bn-\langle \hat{g}^R\rangle_\bn\hat{g}^K-\langle \hat{g}^K\rangle_\bn\hat{g}^A\right).
\end{split}
\en
Similarly, function $\hat{g}^Z$ satisfies the following equation:
\beg\label{Eq4gZ}
\begin{split}
&\left[{\partial}_t+v_F(\bn\cdot\bm{\nabla})\right]\hat{g}^Z+\frac{\lambda_{\text{so}}}{2}\left\{\hat{\bm{\eta}},\bm{\nabla}\hat{g}^Z\right\}\\&+i\Delta_p\left[\hat{\eta}_\bp,\hat{g}^Z\right]=
-i[\phi_1(\br,t_1)-\phi_1(\br,t_2)]\hat{g}^Z\\&-i\phi_2(\br,t_1)\hat{g}^R+i\phi_2(\br,t_2)\hat{g}^A\\&
+\frac{1}{2\tau}\left(\hat{g}^Z\langle \hat{g}^R\rangle_\bn+\hat{g}^A\langle \hat{g}^Z\rangle_\bn-\langle \hat{g}^Z\rangle_\bn\hat{g}^R-\langle \hat{g}^A\rangle_\bn\hat{g}^Z\right).
\end{split}
\en
 By virtue of the constraints imposed by Eq. \eqref{Constraint} only two components of $\hat{g}$ are independent. It is convenient to fix the diagonal components \cite{ANZ2001,LiLi2008}. The extra complication here, in comparison to the procedure carried out in Ref. \cite{LiLi2008}, is that we need to retain the terms up to the first order in powers of $\lambda_{\text{so}}/v_F$:
\beg\label{gRgA}
\begin{split}
\hat{g}^R&=\sqrt{\hat{\mathbbm{1}}-\frac{4\lambda_{\text{so}}}{v_F}\hat{\eta}_\bp-\hat{g}^K\hat{g}^Z}, \\
\hat{g}^A&=-\sqrt{\hat{\mathbbm{1}}-\frac{4\lambda_{\text{so}}}{v_F}\hat{\eta}_\bp-\hat{g}^Z\hat{g}^K}.
\end{split}
\en
The square root should be understood in an operator sense as an expansion in powers of corresponding functions with their products understood as time convolutions and matrix products with respect to spin indices. As for the functions $\hat{g}^K$ and $\hat{g}^Z$, we will look for their expressions by perturbation theory:
\beg\label{Perturb}
\hat{g}^K=\langle\hat{g}^K\rangle_\phi+\delta\hat{g}^K, \quad \hat{g}^Z=\delta\hat{g}^Z.
\en
As we have already mentioned, $\langle\hat{g}^Z\rangle_\phi=0$ by causality. Therefore, up to the linear order in $\delta\hat{g}^{K,Z}$, we have
\beg\label{gRgA2}
\begin{split}
\hat{g}^R&=\hat{\sigma}_0\delta(t_1-t_2)-\frac{2\lambda_{\text{so}}}{v_F}\hat{\eta}_\bp-\langle\hat{g}^K\rangle_\phi\hat{g}^Z, \\
\hat{g}^A&=-\hat{\sigma}_0\delta(t_1-t_2)+\frac{2\lambda_{\text{so}}}{v_F}\hat{\eta}_\bp+\delta\hat{g}^Z\langle\hat{g}^K\rangle_\phi,
\end{split}
\en
where $\hat{\sigma}_0$ is a unit matrix in spin space. These expressions can now be inserted into Eqs. (\ref{Eq4gK}) and (\ref{Eq4gZ}). 
\paragraph{Equation for $\delta\hat{g}^Z$.} In the equation for $\delta\hat{g}^Z$ we only need to keep the terms linear in bosonic fields. It follows then 
\beg\label{Eq4dgZ}
\begin{split}
&\left[\frac{\partial}{\partial t}+v_F(\bn\cdot\bm{\nabla})\right]\delta\hat{g}^Z+i\Delta[\hat{\eta}_\bp,\delta\hat{g}^Z]\\&-\frac{1}{\tau}(\delta\hat{g}^Z-\langle\delta\hat{g}^Z\rangle_\bn)=-2i\phi_2(\br,t_1)\hat{\sigma}_0\delta(t_1-t_2).
\end{split}
\en
Note that here we can safely ignore the gradient term proportional to $\lambda_{\text{so}}$ as it exceeds the accuracy of the calculation. The formal solution of this equation can be written as 
\beg\label{FormalSolve}
\begin{split}
&\delta\hat{g}^Z(t_1,t_2;\bn,\br)=2i\hat{\sigma}_0\delta(t_1-t_2)\\ 
&\times\int d\br'\int dt_3\int\frac{d\bn'}{2\pi}\phi_2(\br',t_3)\Gamma(t_3-t_1;\bn',\bn;\br',\br)
\end{split}
\en
and the kernel of the integral is a diffusion propagator. It satisfies 
\beg\label{Diffuson}
\begin{split}
&(-i\omega+{\mathbf v}_F\cdot\bq)\Gamma(\bn',\bn;\omega,\bq)\\&
+\frac{1}{\tau}\left[\Gamma(\bn',\bn;\omega,\bq)-\langle\Gamma(\bn',\bn;\omega,\bq)\rangle_\bn\right]=
2\pi\delta_{\bn\bn'},
\end{split}
\en
where the Fourier transformation for the diffusion propagator is defined as
\beg\label{DiffusonFour}
\Gamma(t;\bn',\bn;\br,\br')=\int\frac{d\omega d^2\bq}{(2\pi)^3} \Gamma(\bn',\bn;\omega,\bq)e^{i\bq(\br-\br')-i\omega t}.
\en

\paragraph{Equation for $\delta\hat{g}^K$.} To derive an equation for $\delta\hat{g}^K$ we use the same approximations as we did in obtaining Eq. (\ref{Eq4dgZ}). We thus find
\beg\label{Eq4dgK}
\begin{split}
&\left[\frac{\partial}{\partial t}+v_F(\bn\cdot\bm{\nabla})\right]\delta\hat{g}^K+i\Delta[\hat{\eta}_\bp,\delta\hat{g}^K]\\&+
\frac{1}{\tau}\left(\delta\hat{g}^K-\left\langle\delta\hat{g}^K\right\rangle_\bn\right)=2i\phi_2(\br,t_1)\hat{\sigma}_0\delta(t_1-t_2)\\&
-i\left[\phi_1(\br,t_1)-\phi_1(\br,t_1)\right]\langle\delta\hat{g}^K\rangle_\phi\\&+\frac{1}{4\tau}\left[\langle\hat{g}^K\rangle_\phi
\langle\delta \hat{g}^Z\langle\hat{g}^K\rangle_\phi\rangle_\bn-\langle\langle\hat{g}^K\rangle_\phi\rangle_\bn
\delta \hat{g}^Z\langle\hat{g}^K\rangle_\phi\right.\\&\left.
-\langle\hat{g}^K\rangle_\phi\delta \hat{g}^Z\langle\langle\hat{g}^K\rangle_\phi\rangle_\bn+\langle\langle\hat{g}^K\rangle_\phi\delta \hat{g}^Z\rangle_\bn\langle\hat{g}^K\rangle_\phi\right].
\end{split}
\en
In order to write the solution of this equation we separate $\delta\hat{g}^K$ into the (traceless) spin and charge components: $\delta\hat{g}^K=\delta g_0^K+\delta\bm{g}^K\cdot\hat{\bm{\sigma}}$ (we use the same representation for $\hat{g}^K$ as well). In addition, to simplify the analytical analysis of Eq. (\ref{Eq4dgK}), we will assume that $\langle\hat{g}^K\rangle_\phi$ varies slowly on the length scale $L_T=v_F\textrm{min}(1/T,\sqrt{\tau/T})$. Consistent with this assumption is an approximation 
\beg\label{ApproxgK}
\begin{split}
&\langle\hat{g}^K(t_1,t_2;\bn_1,\br_1)\rangle_\phi\approx\langle\langle\hat{g}^K(t_1,t_2;\bn,\br)\rangle_\phi\rangle_\bn\\&+2\bn_1\langle\bn\langle\hat{g}^K(t_1,t_2;\bn,\br)\rangle_\phi\rangle_{\bn}\\&+(\br_1-\br)\bm{\nabla}\langle\langle\hat{g}^K(t_1,t_2;\bn,\br)\rangle_\phi\rangle_\bn.
\end{split}
\en
The expression for the $\delta g_0^K$ is found similarly to the one for $\delta g^Z$. There are three terms in the right hand side of Eq. (\ref{Eq4dgK}), which means that we can look for the solution in the form $\delta g_0^K=\delta g_{0,a}^K+\delta g_{0,b}^K+\delta g_{0,c}^K$. For the first one we find
\beg\label{FormalSolvedg0Ka}
\begin{split}
&\delta{g}_{0,a}^K(t_1,t_2;\bn,\br)=2i\hat{\sigma}_0\delta(t_1-t_2)\\
&\times
\int d\br'\int dt_3\int\frac{d\bn'}{2\pi}\phi_2(\br',t_3)
\Gamma(t_1-t_3;\bn',\bn;\br,\br').
\end{split}
\en
Note the different time dependence in the diffusion propagator, which is due to the opposite sign in front of the third term in left-hand side of Eq. \eqref{Eq4dgK} compared to the similar term in Eq. \eqref{Eq4dgZ}. The second contribution to $\delta g_{0}^K$ originates from the term proportional to $\phi_1$:
\beg\label{FormalSolvedg0Kb}
\begin{split}
&\delta{g}_{0,b}^K(t_1,t_2;\bn,\br)=\\
&-i\int d\br'\int dt_3\int\frac{d\bn'}{2\pi}[\phi_1(\br_1,t_1-t_3)-\phi_1(\br_1,t_2-t_3)]
\\&\times\Gamma(t_3;\bn,\bn';\br,\br')\langle g_0^K(t_1-t_3,t_2-t_3;\bn',\br')\rangle_\phi.
\end{split}
\en
This expression can be further simplified by employing the approximation of Eq. (\ref{ApproxgK}). For the remaining contribution we will provide the approximate expression only neglecting the term which will not contribute to the spin current within the accuracy of our approximations. The expression for $\delta g_{0,c}^K$ we will use in what follows is
\beg\label{FormalSolvedg0Kc}
\begin{split}
&\delta{g}_{0,c}^K(t,t';\bn,\br)\\&\approx\frac{i}{\tau}\prod\limits_{a=2}^4\int dt_a\int d^2\br_2\int\frac{d\bn'}{2\pi}\int\frac{d\bn''}{2\pi}\int\frac{d\bn_2}{2\pi}\\&\times
\left\{\Gamma(t_4-t_3;\bn_2,\bn'';\br_2,\br)-\Gamma(t_4-t_3;\bn_2,\bn';\br_2,\br)\right\}\\&\times\Gamma(t_2;\bn,\bn';\br,\br_1)
\langle\langle g_0^K(t-t_3,t_3;\bn_1,\br_2)\rangle_\phi\rangle_{\bn_1}\\&\times\langle\langle g_0^K(t_3,t'-t_2;\bn_1,\br_2)\rangle_\phi\rangle_{\bn_1}
\phi_2(\br_2,t_4)
\end{split}
\en
and the contribution $\propto\langle\bm{g}^K\rangle\langle\bm{g}^K\rangle$ has been neglected. 

Finally, the expression for $\delta\bm{g}^K$ can be computed in the same way. We will not provide the corresponding expressions here for the same reasons as the approximation we have adopted in the expression for $\delta g_{0,c}^K$: the functions that enter into the expression for 
$\delta\bm{g}^K$ ultimately will not contribute to the spin current.

\paragraph{Equation for $\langle\hat{g}^K\rangle_\phi$.} The expressions for the functions $\delta\hat{g}^Z$ and $\delta\hat{g}^K$ listed above can be used to write the kinetic equation for the function $\langle\hat{g}^K\rangle_\phi$. Averaging both sides over the bosonic fields yields
\beg\label{Eq4gKphi}
\begin{split}
&\left[{\partial}_t+v_F(\bn\cdot\bm{\nabla})\right]\langle\hat{g}^K\rangle_\phi+\frac{\lambda_{\text{so}}}{2}\left\{\hat{\bm{\eta}},\bm{\nabla}\langle\hat{g}^K\rangle_\phi\right\}\\&+i\Delta_p\left[\hat{\eta}_\bp,\langle\hat{g}^K\rangle_\phi\right]=\frac{1}{\tau}\left(\left\langle\langle\hat{g}^K\rangle_\phi\right\rangle_\bn-\langle\hat{g}^K\rangle_\phi\right)\\&-\frac{\lambda_{\text{so}}}{v_F\tau}\left\{\hat{\eta}_{\mathbf{p}},\langle\hat{g}^K\rangle_\phi\right\}+\textrm{St}_{\textrm{el}}\left\{\hat{g}^K\right\}+\textrm{St}_{\textrm{in}}\left\{\hat{g}^K\right\}.
\end{split}
\en
The last two terms on the-right hand side are the collision integrals for the elastic and inelastic scattering processes. 

%$$$$$$$$$$$$$$$$$$$$$$$$$$$$$$$$$$$$$$$$$$
\subsection{Collision integral for the elastic scattering}
%$$$$$$$$$$$$$$$$$$$$$$$$$$$$$$$$$$$$$$$$$$

The elastic scattering contribution to the collision integral is of the form 
\beg\label{Stel1gk}
\begin{split}
&\textrm{St}_{\textrm{el}}\{\hat{g}^K\}=\frac{1}{4\tau}\int dt_3\int\frac{d\bn'}{2\pi}\\&\times\left\{
\hat{F}^R(t_1,t_3;\bn,\bn';\br)\left\langle\hat{g}^K(t_3,t_2;\bn',\br)\right\rangle_\phi\right.\\&\left.-\hat{F}^R(t_1,t_3;\bn',\bn;\br)\left\langle\hat{g}^K(t_3,t_2;\bn,\br)\right\rangle_\phi\right.
\\&\left.+\left\langle\hat{g}^K(t_1,t_3;\bn',\br)\right\rangle_\phi
\hat{F}^A(t_3,t_2;\bn',\bn;\br)\right.\\&\left.-\left\langle\hat{g}^K(t_1,t_3;\bn,\br)\right\rangle_\phi\hat{F}^A(t_3,t_2;\bn,\bn';\br)
\right\}.
\end{split}
\en
Functions $\hat{F}^R$ and $\hat{F}^A$ are related by the equality 
\beg\label{FRFAcc}
\hat{F}^R(\eps,t;\bn_1,\bn_2,\br)=\hat{F}^A(\eps,t;\bn_1,\bn_2,\br)^*.
\en 
For the components of $\hat{F}^{R}$ we have
\beg\label{FR1}
\begin{split}
&\hat{F}^R(t,\eps;\bn_1,\bn_2,\br)\approx i\int\frac{d\omega}{2\pi}\int d^2\br_3\int d^2\br_4\\&\int\frac{d\bn_3}{2\pi}\int\frac{d\bn_4}{2\pi}\mathcal{D}^{R}(t,\omega;\br_3,\br_4)
\left[\Gamma(\omega;\bn_3,\bn_2,\br_3,\br)\right.\\&\left.-\Gamma(\omega;\bn_3,\bn_1,\br_3,\br)\right]\Gamma(\omega;\bn_1,\bn_4,\br,\br_4)
\\&\times
\langle\hat{g}^K(t,\eps-\omega;\bn_4,\br_4)\rangle_\phi.
\end{split}
\en
By construction the elastic collision integral satisfies 
\beg\label{Elastic}
\int\frac{d\bn}{2\pi}\textrm{St}_{\textrm{el}}\left\{\hat{g}^K(t_1,t_2;\bn,\br)\right\}=0,
\en
which means that the total number of electrons on a given energy shell is preserved at any given time $t_1$ and $t_2$.

\paragraph{Expansion in angular components.} 

The terms with the bosonic propagators need to be treated approximately. 
First, we make use of the following approximate relation:
\beg\label{Appro1}
\begin{split}
&\int dt_3\hat{F}^R(t_1,t_3;\bn,\bn';\br)\left\langle\hat{g}^K(t_3,t_2;\bn',\br)\right\rangle_\phi\\&\approx\hat{F}^R(t,\eps;\bn,\bn';\br)\left\langle\hat{g}^K(t,\eps;\bn',\br)\right\rangle_\phi,
\end{split}
\en
where $t=(t_1+t_2)/2$. Here we performed the Fourier transform with respect to $\tau=t_1-t_2$ and the terms proportional to $\partial_t\hat{F}^R(t,\eps;\bn,\bn';\br)$ and $\partial_{\eps}\hat{F}^R(t,\eps;\bn,\bn';\br)$ have been disregarded. Second, only for the collision integral do we adopt an approximation equivalent to an assumption of spatial smoothness:
\beg\label{AngHarm1}
\begin{split}
&\int\frac{d\bn'}{2\pi}\hat{F}^{R}(t,\eps;\bn',\bn,\br)\approx\int\frac{d\bn'}{2\pi}\int\frac{d\bn''}{2\pi}\\&\times\left[1+2(\bn\cdot\bn'')\right]\hat{F}^{R}(t,\eps;\bn',\bn'',\br), \\
&\int\frac{d\bn'}{2\pi}\hat{F}^{A}(t,\eps;\bn,\bn',\br)\approx\int\frac{d\bn'}{2\pi}\int\frac{d\bn''}{2\pi}\\&\times\left[1+2(\bn\cdot\bn'')\right]\hat{F}^{A}(t,\eps;\bn'',\bn',\br).
\end{split}
\en
Furthermore, we use (\ref{ApproxgK}) so that 
\beg\label{AddApp}\nonumber
\begin{split}
&\langle(\bn\cdot\bn'')\hat{F}^{R}(t,\eps;\bn',\bn'',\br)\rangle_{\bn',\bn''}\left\langle\hat{g}^K(t,\eps;\bn,\br)\right\rangle_\phi\\&\approx
\langle(\bn\cdot\bn'')\hat{F}^{R}(t,\eps;\bn',\bn'',\br)\rangle_{\bn',\bn''}\left\langle\langle\hat{g}^K(t,\eps;\bn,\br)\rangle_\phi\right\rangle_\bn. 
\end{split}
\en
 
\paragraph{Interaction correction to the collision integral.} 

Using the approximations above, we find three separate contributions to the collision integral:
\beg
\textrm{St}_{\textrm{el}}\left\{\hat{g}\right\}=\sum\limits_{a=1}^3\textrm{St}^{[a]}_{\textrm{el}}\left\{\hat{g}\right\}.
\en
Below we will write the corresponding expressions for $\textrm{St}^{[a]}_{\textrm{el}}\left\{\hat{g}\right\}$ and will
use the compact notations by replacing $\langle\hat{g}^K(t,\veps;\bn,\br)\rangle_\phi\to\hat{g}(t,\veps;\bn,\br)$. The first one is
\beg\label{Stel0}
\begin{split}
&\textrm{St}^{[1]}_{\textrm{el}}\left\{\hat{g}\right\}=-\frac{2}{\tau}\int\frac{d\omega}{2\pi}n_i{\cal R}_{ij}^{[1]}(\omega)\\&\times
\left\langle g_0(t,\veps-\omega;\bn,\br)\right\rangle_\bn\left\langle n_j\hat{g}(t,\veps;\bn,\br)\right\rangle_\bn, 
\end{split}
\en
where the kernel of the integral is 
\beg\label{R1}
\begin{split}
&{\cal R}_{ij}^{[1]}(\omega)=\textrm{Im}\int\frac{d^2\bq}{(2\pi)^2}\mathcal{D}^R(\omega,\bq)\\&\times\left[\langle\Gamma n_i\rangle\langle\Gamma n_j\rangle-\frac{\delta_{ij}}{2}\left(\langle\Gamma\rangle\langle\Gamma\rangle-\langle\Gamma\Gamma\rangle\right)\right].
\end{split}
\en
In the expression above the angular averages of the diffusion propagators are defined as
\beg\label{AvDiffs}
\begin{split}
\langle f\Gamma g\rangle&=\int\frac{d\bn d\bn'}{(2\pi)^2}f(\bn)\Gamma(\bn,\bn';\omega,\bq)g(\bn'), \\
\langle f\Gamma g\Gamma h\rangle&=\int\frac{d\bn d\bn' d\bn''}{(2\pi)^3}f(\bn)\Gamma(\bn,\bn';\omega,\bq)
\\&\times g(\bn')\Gamma(\bn',\bn'';\omega,\bq)h(\bn'').
\end{split}
\en
The second contribution to the collision integral is 
\beg\label{Stel2}
\begin{split}
\textrm{St}^{[2]}_{\textrm{el}}\left\{\hat{g}\right\}&=\frac{i}{\tau}\int\frac{d\omega}{2\pi}n_i\overline{\cal R}_{ij}^{[2]}(\omega)\\&\left\langle n_j\hat{g}(t,\veps-\omega;\bn,\br)\right\rangle_\bn
\left\langle \hat{g}(t,\veps;\bn,\br)\right\rangle_\bn\\&-\frac{i}{\tau}\int\frac{d\omega}{2\pi}n_i\left[\overline{\cal R}_{ij}^{[2]}(\omega)\right]^*\\&\times\left\langle \hat{g}(t,\veps;\bn,\br)\right\rangle_\bn\left\langle n_j\hat{g}(t,\veps-\omega;\bn,\br)\right\rangle_\bn, \\
\end{split}
\en
and the kernel of the integral is 
\begin{align}\label{R2}
&\overline{\cal R}_{ij}^{[2]}(\omega)=\int\frac{d^2\bq}{(2\pi)^2}\mathcal{D}^R(\omega,\bq)\nonumber \\ 
&\times\left(\langle\Gamma\rangle\langle n_i\Gamma n_j\rangle-\langle\Gamma n_i\rangle\langle\Gamma n_j\rangle-\langle\Gamma n_i\Gamma n_j\rangle\right).
\end{align}
Finally, the third contribution to the collision integral appears as a result of the gradient expansion of Eq. (\ref{ApproxgK}) in the expressions for $\hat{F}^{R,A}$.
We find the following expression for it: 
\beg\label{Stelgrad}
\begin{split}
\textrm{St}^{[3]}_{\textrm{el}}\left\{\hat{g}\right\}&=-\frac{2}{\tau}\int\frac{d\omega}{2\pi}n_i{\cal R}_{i\alpha}^{[3]}(\omega)\\
&\times\nabla_\alpha\left\langle g_0(t,\veps-\omega;\bn,\br)\right\rangle_\bn
\left\langle \hat{g}(t,\veps;\bn,\br)\right\rangle_\bn, \\
\end{split}
\en
where
\begin{align}
&{\cal R}_{i\alpha}^{[3]}(\omega)=-\textrm{Re}\int\frac{d^2\bq}{(2\pi)^2}\mathcal{D}^R(\omega,\bq)\nonumber \\ 
&\times\left(\langle\Gamma\rangle\frac{\partial}{\partial q_\alpha}\langle n_i\Gamma\rangle-\langle\Gamma n_i\frac{\partial}{\partial q_\alpha}\Gamma\rangle-\langle\Gamma n_i\rangle\frac{\partial}{\partial q_\alpha}\langle\Gamma\rangle\right).
\end{align}
We discuss the specific details pertaining to the dependence of the kernel functions on frequency below in the section devoted to the calculation of the spin- current. In addition, it should be noted that in the presence of external field, the vector potential enters spatial gradients in the gauge invariant way, therefore inclusion of the field amounts to the replacement, e.g., $\bm{\nabla}\to\bm{\partial}=\bm{\nabla}+e\bm{\mathcal{E}}\partial_\epsilon$. 

%$$$$$$$$$$$$$$$$$$$$$$$$$$$$$$$$$$$$$$$$$$$$
\subsection{Collision integral for the inelastic scattering}
%$$$$$$$$$$$$$$$$$$$$$$$$$$$$$$$$$$$$$$$$$$$$

Formally, the inelastic scattering collisions are accounted for by the collision integral 
\beg\nonumber
\textrm{St}_{\textrm{in}}\{\hat{g}^K\}=-i\left\langle\left[\phi_1(\br,t_1)-\phi_1(\br,t_2)\right]\delta\hat{g}^K\right\rangle_\phi.
\en
From this expression it is clear that $\textrm{St}_{\textrm{in}}\{\hat{g}^K\}(t_1,t_1,\bn,\br)=0$. This means that the total number of particles moving in a given direction $\bn$ is conserved. We are not going to give a detailed expression for $\textrm{St}_{\textrm{in}}\{\hat{g}^K\}$ since in what follows we
are going to limit our analysis to the limit of small electric fields, which means we will retain the terms linear in electric field. In this approximation inelastic processes do not affect the $\langle\hat{g}\rangle_\bn$.

%#############################################################################################################################
%#############################################################################################################################
%#############################################################################################################################
%#############################################################################################################################
%#############################################################################################################################

\section{Equation for the density matrix}\label{sec:DensityMatrix}
The analysis of preceding sections prepared us to derive the diffusion equation for the density matrix 
$\hat{\rho}_{\veps}=\langle\hat{g}(t,\veps;\bn,\br)\rangle_\bn$ in the presence of the finite electric field $\bm{\mathcal{E}}$. Using the result of our earlier discussion our starting point is the kinetic equation 
\beg\label{Eilen2}
\begin{split}
&\left({\partial}_t+v_F\bn\cdot \bm{\partial}\right)\hat{g}+\frac{\lambda_{\text{so}}}{2}\left\{\hat{\bm{\eta}},\bm{\partial}\hat{g}\right\}+i\Delta\left[\hat{\eta}_\bp,\hat{g}\right]\\&=\frac{\left\langle\hat{g}\right\rangle_\bn-
\hat{g}}{\tau}-\frac{\lambda_{\text{so}}}{v_F\tau}\left\{\hat{\eta}_\bp,\left\langle\hat{g}\right\rangle_\bn\right\}+\textrm{St}_{\textrm{el}}\{\hat{g}\},
\end{split}
\en
where now $\Delta=\lambda_{\text{so}}p_F$. It is clear that if we were to average both sides of this equation over $\bn$, then the matrix function $2\bn\langle\bn'\hat{g}\rangle_{\bn'}$ will also appear. If we were interested in the spin dynamics we could have determined $2\bn\langle\bn'\hat{g}\rangle_{\bn'}$ perturbatively and thus established the corresponding correction to $\langle \hat{g}\rangle_\bn$ \cite{LiLi2008}.  
This procedure would also work for the computing Al'tshuler-Aronov correction to conductivity; however, it is not sufficient for the spin Hall effect.  

In order to obtain the equation for the density matrix, which would ultimately produce a reactive contribution to the spin current, we first rewrite Eq. (\ref{Eilen2}) 
as follows:
\beg\label{Eq2Iter}
\begin{split}
\left({\partial}_t+\frac{1}{\tau}\right)\hat{g}+i\Delta\left[\hat{\eta}_\bp,\hat{g}\right]=\hat{K}[\hat{g}],
\end{split}
\en
where we use $\hat{K}[\hat{g}]=\hat{K}_0[\hat{\rho}_\veps]+\hat{K}_1[\hat{g}]
+\hat{K}_2[\hat{\cal W},\hat{g}]$ and
\beg\label{Kis}
\begin{split}
&\hat{K}_0[\hat{\rho}_\veps]=\frac{\hat{\rho}_\veps}{\tau}-\frac{\lambda_{\text{so}}}{v_F\tau}\left\{\hat{\eta}_\bp,\hat{\rho}_\veps\right\}, \\
&\hat{K}_1[\hat{g}]=-v_F\left(\bn\cdot\bm{\partial}\right)\hat{g}-\frac{\lambda_{\text{so}}}{2}\left\{\hat{\bm{\eta}},\bm{\partial}\hat{g}\right\}, \\
&\hat{K}_2[\hat{g}]=\textrm{St}_{\textrm{el}}\{\hat{g}\}.
\end{split}
\en
The formal solution of Eq. (\ref{Eq2Iter}) is obtained by using the Fourier transform with respect to time $t$:
\beg\label{Solve4g}
\begin{split}
\hat{g}&=\frac{(Z_\omega^2+2\Delta^2)}{Z_\omega(Z_\omega^2+4\Delta^2)}\hat{K}+\frac{2\Delta^2}{Z_\omega(Z_\omega^2+4\Delta^2)}\hat{\eta}_\bp\hat{K}\hat{\eta}_\bp\\&-\frac{i\Delta}{Z_\omega^2+4\Delta^2}\left[\hat{\eta}_\bp,\hat{K}\right]\equiv{\cal F}[\hat{K}].
\end{split}
\en
Here $Z_\omega=-i\omega+\tau^{-1}$. We can now derive the equation for $\hat{\rho}_\veps$ by solving Eq. (\ref{Solve4g}) by consecutive iterations in analogy to the approach used earlier in Ref. \cite{MSH2004}. 
 
\paragraph{First iteration.} 
To find the first iterative solution, $\hat{g}^{(0)}$, in Eq. (\ref{Solve4g}) we replace $\hat{K}$ with $\hat{K}_0$, and average both sides of the equation over $\bn$: 
\beg\label{ZerothDetails}
\langle\hat{g}^{(0)}\rangle_\bn=\left\langle{\cal F}[\hat{K}_0]\right\rangle_\bn.
\en
As a result, we find 
\beg\label{rho0w0}
\begin{split}
-i\omega \hat{\rho}_\veps&=-\frac{\hat{\rho}_\veps}{\tau_s}+\frac{\hat{\bm{\eta}}\hat{\rho}_\veps\hat{\bm{\eta}}}{2\tau_s}.
\end{split}
\en
Here we introduced the spin-relaxation time $\tau_s$ expressed in terms of a dimensionless parameter $\zeta$, 
\begin{equation}
\tau_s=[2\zeta^2/({1+4\zeta^2})]^{-1}\tau, \quad \zeta=\tau\Delta.
\end{equation}
The expression for the matrix function $\hat{\cal W}$ is found by multiplying $\hat{g}^{(0)}$ by $\bn$ and performing the averaging over $\bn$. It follows 
\beg\label{W0}
2\bn\langle\bn'\hat{g}^{(0)}\rangle_{\bn'}=-\frac{i\zeta}{1+4\zeta^2}\left[\hat{\eta}_\bp,\hat{\rho}_\veps\right]-\frac{\lambda_{\text{so}}}{v_F}\left\{\hat{\eta}_\bp,\hat{\rho}_\veps\right\}. 
\en

\paragraph{Second iteration.} 
The second iteration leads to the appearance of the linear-in-gradient terms in the equation for the density matrix as well as in the expression for $2\bn\langle\bn'\hat{g}\rangle_{\bn'}$. By definition
\beg\label{g1}
\langle\hat{g}^{(1)}\rangle_\bn=\left\langle{\cal F}[\hat{K}_1(\hat{g}^{(0)})]\right\rangle_\bn.
\en
After a somewhat lengthy calculation (the details can be found in the Appendix A) we find
\beg\label{Eq4rho1}
-i\omega\hat{\rho}_\veps=-\frac{\hat{\rho}_\veps}{\tau_s}+\frac{\hat{\bm{\eta}}\hat{\rho}_\veps\hat{\bm{\eta}}}{2\tau_s}+B\left\{\hat{\bm{\eta}},\bm{\partial}\hat{\rho}_\veps\right\}+iC\left[\hat{\bm{\eta}},\bm{\partial}\hat{\rho}_\veps\right],
\en
where the expressions for the coefficients $B$ and $C$ are given by
\beg\label{Coeffs}
B=\frac{\lambda_{\text{so}}\zeta^2}{1+4\zeta^2}, \quad C=\frac{v_F\zeta}{(1+4\zeta^2)^2}.
\en
In order to calculate the gradient correction to $\hat{\cal W}$, it will be sufficient to use the approximation 
\beg\label{apprho}
\bm{\partial}\hat{\rho}_{\veps}({\mathbf x})\approx\frac{1}{2}\bm{\partial}f_\veps\hat{\sigma}_0,
\en
where function $f_\veps$ is the single particle density. We thus have
\beg\label{W1}
\begin{split}
2\bn\langle\bn'\hat{g}^{(1)}\rangle_{\bn'}&=-\frac{v_F\tau}{2}\left\{\left(\bn\cdot\bm{\partial}{f}_\veps\right)\hat{\sigma}_0\phantom{\frac{}{}}\right.\\&\left.+\left(\frac{\lambda_{\text{so}}}{v_F}\right)^2\frac{i\zeta}{(1+4\zeta^2)}\left[\hat{\eta}_\bp,\hat{\bm{\eta}}\cdot\bm{\partial}{f}_\veps\right]\right\}.
\end{split}
\en
We can now use these expressions to compute the interaction correction to the equation for the density matrix. 

\paragraph{Third iteration.} 
In our case, the third iteration produces terms still linear in gradient, but (by construction) they have additional smallness due to interaction effects. Thus, it is defined as
\beg\label{g2}
\langle\hat{g}^{(2)}\rangle_\bn=\left\langle{\cal F}[\hat{K}_2(\hat{g}^{(1)})]\right\rangle_\bn.
\en
The details of the calculation are given in Appendix B, so here we give the final form of the equation for the density matrix:
\beg\label{Interact}
\begin{split}
-i\omega\hat{\rho}_\veps&=-\frac{\hat{\rho}_\veps}{\tau_s}+\frac{\hat{\bm{\eta}}\hat{\rho}_\veps\hat{\bm{\eta}}}{2\tau_s}+B\left\{\hat{\bm{\eta}},\bm{\partial}\hat{\rho}_\veps\right\}+iC\left[\hat{\bm{\eta}},\bm{\partial}\hat{\rho}_\veps\right]\\&
-eI_{ij}\{f_\veps\}[\bm{e}_z\times\bm{\mathcal{E}}]_i\hat{\sigma}_j,
\end{split}
\en
where in the last term we have used Eq. (\ref{apprho}) and the summation over repeated indices is assumed. The matrix elements of $I_{ij}(\veps)$ are
\beg\label{Iij}
\begin{split}
I_{ij}[f_\veps]&=\frac{\lambda_{\text{so}}^2\zeta C}{2v_F^2}\int \frac{d\omega}{2\pi}\left[{\cal R}_{ij}^{[0]}(\omega)\left(\frac{\partial f_{\veps-\omega}}{\partial \veps}\right)f_\veps\right.\\&\left.+\left({\cal R}_{ij}^{[1]}(\omega)+\zeta {\cal R}_{ij}^{[2]}(\omega)\right)f_{\veps-\omega}\frac{\partial f_\veps}{\partial\veps}\right]
\end{split}
\en
and we introduced 
\beg\label{R0R2Again}
{\cal R}_{ij}^{(0)}(\omega)=\textrm{Im}[\overline{\cal R}_{ij}^{(2)}(\omega)], ~{\cal R}_{ij}^{(2)}(\omega)=-\textrm{Re}[\overline{\cal R}_{ij}^{(2)}(\omega)].
\en
We solve the Eq. (\ref{Interact}) up to the linear order in electric field. To do that, we fist use the spin-basis representation for the density matrix:
\beg\label{rhoepsexpand}
\hat{\rho}_\veps=\frac{f_\veps}{2}\hat{\sigma}_0+(\bm{s}_\veps\cdot\hat{\bm{\sigma}}).
\en
Then, for the components of the spin-density $\bm{s}_\veps$ in the bulk of the sample we find
\beg\label{sepsbulk}
\begin{split}
s_\veps^i&=-e\tau_s\left(B\frac{\partial f_\veps}{\partial\veps}\delta_{ij}+I_{ij}\{f_\veps\}\right)[\bm{e}_z\times\bm{\mathcal{E}}]_j.
\end{split}
\en
Three comments are in order: (1) the interaction correction to the spin-density is small, $O(\lambda_{SO}^2/v_F^2)$; (2) it is proportional to the coefficient $C$, while the first term is proportional to the coefficient $B$; and (3) since electric field lies in the plane, the $z$ component of the spin polarization is not generated even in the presence of interactions. As we will see below, the magnitude of the reactive part of the spin Hall conductivity is determined by the value of $B$. Since we assumed that the coefficient $\zeta$ can be arbitrary, it follows that in the limit when $\zeta\gg 1$, the interaction correction acquires an extra small prefactor $O(1/\zeta^2)$.

%#############################################################################################################################
%#############################################################################################################################
%#############################################################################################################################
%#############################################################################################################################
%#############################################################################################################################

\section{Spin-current}\label{sec:SpinCurrent}
We use the standard (i.e. conventional) definition of the spin current defined by a symmetric product of the spin and velocity operators, 
$\hat{\cal J}_k^i=(\hat{\sigma}_i\hat{v}_k+\hat{v}_k\hat{\sigma}_i)/4$. Alternatively, one may also consider the definition in which the spin current satisfies the continuity equation \cite{Sugimoto2006}. Our subsequent results concerning the interaction correction to spin Hall conductivity are not affected by this choice.

The expectation value of the spin current $\hat{\cal J}_k^i$, as defined above, can be expressed in terms of the Keldysh component of the Green's function:
\beg\label{SpinCurrent}
{\cal J}_k^i=\frac{1}{8m}\textrm{Tr}\left\{\hat{\sigma}^i\left(\nabla_k'-\nabla_k\right)\hat{G}^K(x,x')\right\}_{{x}'\to{x}}+\frac{\lambda_{\text{so}}}{2}\eps^{ikz}N.
\en
Here $N$ is the total particle number
\beg\label{PNumber}
N=\nu_F\int\limits_{-\infty}^\infty d\veps\textrm{Tr}(\hat{\rho}_\veps),
\en
$\eps^{ikz}$ is an absolutely anti-symmetric tensor of the second rank and $\nu_F=m/2\pi$ is the density of states at the Fermi level. The Dyson equation (\ref{Dyson4Ferms}) can be manipulated in a way which allows us to express $\hat{G}^K$ in terms of $\hat{\rho}_\veps$ via $\hat{G}^K=\hat{G}^R\hat{\Sigma}\hat{G}^A$ (as before, this product should be understood as convolution in time and space and matrix product for spin). After performing the Fourier transformations and making use of Eq. (\ref{SelfEnergy}) we find
\beg\label{JikFull}
\begin{split}
&{\cal J}_k^i=\frac{i}{8\pi m\tau}\left(\nabla_k'-\nabla_k\right)\int\limits_{-\infty}^\infty d\veps\int d^2{\mathbf y}\\
&\times\textrm{Tr}\left\{\hat{\sigma}_i\hat{G}_{\veps}^R({\mathbf x}-{\mathbf y})\hat{\rho}_\veps({\mathbf y})\hat{G}_{\veps}^A({\mathbf y}-{\mathbf x}')\right\}_{{\mathbf x}'\to{\mathbf x}}\!\!+\frac{\lambda_{\text{so}}}{2}\eps^{ikz}N.
\end{split}
\en
To streamline the calculation of the area integral we will again rely on the gradient expansion for the density matrix:
\beg\label{rhograd}
\begin{split}
\hat{\rho}_\veps({\mathbf y})&\approx\hat{\rho}_\veps({\mathbf x})+\left({\mathbf y}-{\mathbf x}\right)\cdot\bm{\partial}\hat{\rho}_{\veps}({\mathbf x}).
\end{split}
\en
As the next step, the integral over the area can be done by going into the momentum representation. We use the following expressions for the retarded and advanced components of the Green's function: 
\beg\label{GRA}
\hat{G}^{R(A)}(\veps,\bp)=\frac{1}{2}\sum\limits_{a=\pm}\frac{\hat{\sigma}_0-\textrm{sgn}(a)\hat{\eta}_\bp}{\veps-\frac{p^2}{2m}+E_F+
\textrm{sgn}(a)\lambda_{\text{so}}p\pm\frac{i}{2\tau}}.
\en
Given Eq. (\ref{rhograd}), it is clear that there are two contributions to the spin current stemming from Eq. (\ref{JikFull}) 
\beg\label{TwoContribs}
{\cal J}_k^i={j}_{k}^i+\tilde{j}_k^i.
\en
The first contribution originates from the $\hat{\rho}_\veps(\bx)$, and it is found to be 
\beg\label{FirstOne}
{j}_{k}^i=\frac{v_F\zeta}{1+4\zeta^2}\left(\delta_z^i S^k-\delta_k^i S^z\right),
\en
where $S^k$ is the $k$th component of the spin polarization
\beg\label{SpinPol}
\bm{S}=\frac{\nu_F}{2}\int\limits_{-\infty}^\infty d\veps\textrm{Tr}\left(\hat{\bm{\sigma}}\hat{\rho}_\veps\right).
\en 
Equation (\ref{FirstOne}) represents a reactive (dissipationless) part of the spin current. 
The second contribution to the spin-current is proportional to $\bm{\partial}\hat{\rho}_{\veps}({\mathbf x})$. Since this term is small, we approximate the gradient term as in Eq. (\ref{apprho}).
Using the integration by parts it turns out that the only nonvanishing contribution to the current is proportional to $\partial_{p_j}\hat{G}_{\bp\veps}^R$ and is given by 
\beg\label{SecondOne}
\tilde{j}_k^i=-\left(\frac{e}{2\pi}\right)\frac{\zeta^2}{1+4\zeta^2}[\bm{e}_z\times\bm{\mathcal{E}}]_k\delta_z^i.
\en
This contribution can be interpreted as Drude (dissipative) part of the spin current. 

%$$$$$$$$$$$$$$$$$$$$$$$$$$$$$$$$$$$$$$$$
\subsection{Noninteracting case} 
%$$$$$$$$$$$$$$$$$$$$$$$$$$$$$$$$$$$$$$$$

Let us briefly discuss the noninteracting case, $I_{ij}\{n_\veps\}=0$ and assuming the static limit. For the reactive part of the spin current from Eqs. 
(\ref{sepsbulk}) and (\ref{SpinPol}) we find
\beg\label{jik0}
\begin{split}
{j}_{k}^i&=\delta_z^i\frac{v_F\zeta}{1+4\zeta^2}\left(\nu_Fe\tau\lambda_{\text{so}}\right)[\bm{e}_z\times\bm{\mathcal{E}}]_k\\&=\left(\frac{e}{2\pi}\right)
\frac{\delta_z^i\zeta^2}{1+4\zeta^2}[\bm{e}_z\times\bm{\mathcal{E}}]_k=-\tilde{j}_k^i,
\end{split}
\en
In deriving this expression, we took into account that for small deviations from equilibrium we can express $f_\veps$ in terms of the Fermi distribution functions $n_F(\veps)=[\exp(\veps/T)+1]^{-1}$ and $T$ is temperature. Specifically, given definitions (\ref{rhoepsexpand} and \ref{PNumber}) it follows that $f_\veps=n_F(\veps+\Delta)+n_F(\veps-\Delta)$, so that 
\beg\label{Intdfveps}
\int\limits_{-\infty}^{+\infty}d\veps\frac{\partial f_\veps}{\partial\veps}=-2.
\en
Thus, we verify that in the absence of interactions the reactive and dissipative contributions to the spin current cancel each other out in the bulk.

%$$$$$$$$$$$$$$$$$$$$$$$$$$$$$$$$$$$$$$$$
\subsection{Finite frequency spin-Hall response} 
%$$$$$$$$$$$$$$$$$$$$$$$$$$$$$$$$$$$$$$$$

It is of interest to consider spin Hall effect in response to the alternating field $\bm{\mathcal{E}}(t)=\bm{\mathcal{E}}_\omega\cos(\omega t)$. After the Fourier transformation, the equation for the density matrix can be found in the following form: 
\beg
-i\omega\hat{\rho}_\veps=-\frac{\hat{\rho}_\eps}{\tau_{s\omega}}+\frac{\hat{\bm{\eta}}\hat{\rho}_\eps\hat{\bm{\eta}}}{2\tau_{s\omega}}+eB_\omega[\bm{e}_z\times\hat{\bm{\sigma}}]\bm{\mathcal{E}}_\omega\frac{\partial f}{\partial\eps}
\en
where 
\begin{equation}
\frac{1}{\tau_{s\omega}}=\frac{2\Delta^2\tau}{(1-i\omega\tau)^2+4\zeta^2},\quad B_\omega=\frac{\lambda_{\text{so}}\zeta^2}{(1-i\omega\tau)^2+4\zeta^2}.
\end{equation}
With the parametrization from Eq. \eqref{rhoepsexpand} the equation for the spin component becomes 
\begin{equation}\label{eq:s-omega}
(\tau^{-1}_{s\omega}-i\omega)\bm{s}_\veps=-eB_\omega[\bm{e}_z\times\bm{\mathcal{E}}_\omega]\frac{\partial f}{\partial\eps}.
\end{equation} 
The expression for the spin current has been derived earlier and can be reduced to 
\begin{align}
j^l_k(\omega)&=\delta^l_z\left(-\frac{e}{2\pi}\right)\frac{\Delta^2}{Z^2_\omega+4\Delta^2}[\bm{e}_z\times\bm{\mathcal{E}}_\omega]_k\nonumber \\ 
&+\delta^l_z\frac{v_F\Delta}{\tau(Z^2_\omega+4\Delta^2)}S^k(\omega),\quad \bm{S}=\nu_F\int d\veps\bm{s}_\veps
\end{align}
We solve for $\bm{s}_\veps$ from Eq. \eqref{eq:s-omega}, and then after the final integration and a few simple algebra steps extract the spin Hall conductivity 
\begin{equation}
\sigma_{\text{sH}}(\omega)=\left(\frac{e\zeta^2}{2\pi}\right)\frac{\omega\tau}{\omega\tau[4\zeta^2-(i+\omega\tau)^2]+2i\zeta^2}.
\end{equation}
This result was plotted in Fig. \ref{Fig-ReIm-sigma} for various values of $\zeta$. In the ballistic limit $\zeta\gg1$ one can easily extract the limiting cases 
\begin{equation}
\sigma_{\text{sH}}(\omega)=\frac{e}{2\pi}\times\left\{\begin{array}{cc}-i\omega\tau & \omega\ll\tau^{-1}_s \\ 1/4 & \tau^{-1}_s\ll\omega\ll\tau^{-1} \\ 
-\zeta^2/(\omega\tau)^2 & \omega\gg\tau^{-1}\end{array}\right.
\end{equation}
In contrast, in the diffusive limit $\zeta\ll1$, the maximum value of the spin Hall conductivity remains strongly suppressed 
$\sigma_{\text{sH}}\approx e\zeta^2/(2\pi)$. These results are consistent with earlier calculations reported in Refs. \cite{MSH2004,Chalaev2005}.  

%$$$$$$$$$$$$$$$$$$$$$$$$$$$$$$$$$$$$$$$$
\subsection{Interaction correction to spin-current} 
%$$$$$$$$$$$$$$$$$$$$$$$$$$$$$$$$$$$$$$$$

In order to find the interaction correction to spin current, we first need to compute the spin polarization (\ref{SpinPol}) by integrating Eq. (\ref{Iij}) over $\veps$ using our result Eq. (\ref{sepsbulk}). Inserting the resulting expression for $S^k$ into Eq. (\ref{FirstOne}) and given the cancellation of the noninteracting terms (\ref{jik0}), the net spin current is given by the sum of two terms, 
\beg\label{TotalCurrent}
{\cal J}_k^i=\delta_1J_k^i+\delta_2 J_k^i.
\en
For the first one we utilize the property of the kernel functions 
\beg\label{R01prop}
\int\limits_{-\infty}^\infty\frac{d\omega}{2\pi}{\cal R}_{ij}^{[0]}(\omega)=\int\limits_{-\infty}^\infty\frac{d\omega}{2\pi}{\cal R}_{ij}^{[1]}(\omega)=0 
\en
along with the fact that ${\cal R}_{ij}^{[0]}(\omega)=-{\cal R}_{ij}^{[1]}(\omega)=\delta_{ij}{\cal R}(\omega)$ (see Appendix C for details). It follows then 
\beg\label{FinalJik}
\begin{split}
&\delta_1{J}_k^i=-\delta_z^i\left(\frac{e}{8\pi}\right)\left(\frac{\zeta}{1+4\zeta^2}\right)^2\frac{\lambda_{\text{so}}}{v_F}
\\&\times\int\frac{d\omega}{2\pi}\sum\limits_{ab}\frac{\partial}{\partial\omega}\left[{\omega_{ab}}
\coth\left(\frac{\omega_{ab}}{2T}\right)\right]{\cal R}(\omega)[\bm{e}_z\times\bm{\mathcal{E}}]_k.
\end{split}
\en
In this expression the summation is performed over $(a,b)=\pm1$, $\omega_{ab}=\omega+\textrm{sgn}(a-b)\lambda_{\text{so}}p_F$ and we used the following identity:
\beg\label{Identities}
\begin{split}
&\int\limits_{-\infty}^\infty{d\veps}\frac{\partial f_{\veps-\omega}}{\partial\veps}f_{\veps}=-2+\sum\limits_{ab}\frac{\partial}{\partial\omega}\left[\frac{\omega_{ab}}{2}
\textrm{coth}\left(\frac{\omega_{ab}}{2T}\right)\right].
\end{split}
\en
As could have been expected, the magnitude of the spin-current correction is determined by the additional smallness of the dimensionless parameter $\lambda_{\text{so}}/v_F$. 
For the remaining contribution we found 
\beg\label{RemainJik}
\begin{split}
\delta_2 J_k^i&=-\delta_z^i\left(\frac{e}{8\pi}\right)\left(\frac{\zeta}{1+4\zeta^2}\right)^2\frac{\lambda_{\text{so}}\zeta}{2v_F}
\\&\times\int\frac{d\veps d\omega}{2\pi}{\cal R}_2(\omega)f_{\veps-\omega}\frac{\partial f_\veps}{\partial\veps}
[\bm{e}_z\times\bm{\mathcal{E}}]_k, 
\end{split}
\en
where we again used ${\cal R}_{ij}^{[2]}(\omega)=\delta_{ij}{\cal R}_2(\omega)$. The remaining integrals with the collision kernels are known to be divergent in the ultraviolet and thus require regularization. The same issue arises in the context of interaction corrections to the conductivity (see Ref. \cite{ANZ2001} for a detailed discussion) since the corresponding collision integrals have exactly the same physical origin. We thus adopt the same procedure and replace 
\begin{equation}
\int\limits^{\infty}_{0}d\omega\frac{\partial}{\partial\omega}\left(\omega\coth\frac{\omega}{2T}\right)\to-2T+E_F\coth\frac{E_F}{2T}
\end{equation}
where $E_F$ is put for the upper limit of the integral. This is legitimate and consistent with the approximations in momentum integration, where one relies on fast convergence in order to set the integration limit (otherwise determined by the Fermi energy) to infinity and to set all momenta in the numerator to the Fermi momentum in magnitude. This approach enables us to find the low-temperature asymptote since we are interested only in temperatures $T\ll E_F$, where the second term is essentially a temperature independent constant. These considerations lead to the following estimate in the ballistic limit 
\begin{equation}
\delta\sigma_{\text{sH}}\simeq\frac{e}{p_Fl}\left(\frac{T}{E_F}\right)\ln\left(\frac{E_F}{T}\right),
\end{equation}
where we used ${\cal R}_2(T\tau\gg 1)\approx\frac{1}{2E_F}\ln\left(\frac{v_Fq^*}{|\omega|}\right)$ with the momentum cutoff $q^*\approx p_F$. Curiously, we notice that spin-orbit coupling $\lambda_{\text{so}}$ drops out from this result.  We have not attempted to further verify cancellations of interaction corrections in the static limit stemming for all the interaction channels, but based on the general grounds we conjecture that this cancellation indeed takes place. We remind that cancellation of this kind were demonstrated earlier diagrammatically in the context of AHE \cite{Langenfeld1991,Muttalib2007,Li2020}.  The scale of $\delta\sigma_{\text{sH}}$ gives an order of magnitude estimate for the quantum interference effects to the spin Hall conductivity in the range where it is frequency independent $\omega\tau_s>1$. As expected, this correction contains an extra smallness in $1/(p_Fl)$. In the diffusive case further suppressions occur in the parameter of $\zeta\ll1$.       

%$$$$$$$$$$$$$$$
\section{Summary}
%$$$$$$$$$$$$$$$

To summarize, in this work we have attempted to consistently describe the effect of impurity scattering and electron-electron interaction on the spin Hall conductivity of two-dimensional electron gas in the Rashba model. Our approach is valid for the arbitrary relationship between temperature and elastic scattering time and as such covers both ballistic region, $T\tau>1$, and diffusive transport regime, $T\tau<1$. We found that interactions lead to the temperature-dependent correction to the spin Hall conductivity. We also provided a detailed discussion of the frequency dependence of spin Hall effect. On a technical level, our approach is distinct from the previously employed methods, but it is complementary in many ways, as we reproduced various results in the course of derivation. This technique can be successfully applied to analyze other relevant models and problems. For instance, the calculation can be pushed further to derive the spin- and charge-diffusion equations from which the spin-relaxation time can be determined explicitly. With the derived collision kernels it is possible to elucidate 
the influence of the electron-electron interaction on the dynamics of spin relaxation. Future work may focus on extensions of the theory to hydrodynamic regime of strong electron interactions where conserved spin currents may lead to peculiar magnetotransport phenomena.    

%#############################################################################################################################
%#############################################################################################################################
%#############################################################################################################################
%#############################################################################################################################
%#############################################################################################################################

\section*{Acknowledgments}
We thank Maxim Khodas for the illuminating discussions and detailed explanations related to the results of Ref. \cite{Khodas2005}.  
This work was financially supported by the National Science Foundation Grants No. DMR-2002795 (M. D.) and DMR-2203411 (A. L.). 

%#############################################################################################################################
%#############################################################################################################################
%#############################################################################################################################
%#############################################################################################################################
%#############################################################################################################################

\begin{appendix}

%$$$$$$$$$$$$$$$$$$$$$$$$$$$$$$$$$$$$$$$$$$$$$$$$$$$$$$$$$$$$$$$$$$$$$$$$$$$
\section{Derivation of the expressions for the coefficients in the equation for the density matrix}
%$$$$$$$$$$$$$$$$$$$$$$$$$$$$$$$$$$$$$$$$$$$$$$$$$$$$$$$$$$$$$$$$$$$$$$$$$$$

In this section we provide additional details for the calculation of the coefficients $B$ and $C$ in Eqs. (\ref{Coeffs}). 
For convenience we will consider the separate contributions as $\hat{K}_1=\hat{K}_1^{[a]}+\hat{K}_1^{[b]}$ with
\beg\label{Separ}
\begin{split}
\hat{K}_1^{[a]}[\hat{g}^{(0)}]&=-v_F\left(\bn\cdot\bm{\partial}\right)\hat{g}^{(0)},  \\
\hat{K}_1^{[b]}[\hat{g}^{(0)}]&=-\frac{\lambda_{\text{so}}}{2}\left\{\hat{\bm{\eta}},\bm{\partial}\hat{g}^{(0)}\right\}.
\end{split}
\en
A simple calculation of angular averages yields
\beg\label{AvK1ab}
\begin{split}
&\left\langle\hat{K}_1^{[a]}\right\rangle_\bn=\frac{iv_F\Delta}{2(Z_\omega^2+4\Delta^2)\tau}\left[\hat{\bm{\eta}},\bm{\partial}\hat{\rho}_\veps\right]+\frac{\lambda_{\text{so}}}{2Z_\omega\tau}\left\{\hat{\bm{\eta}},\bm{\partial}\hat{\rho}_\veps\right\}, \\
&\left\langle\hat{K}_1^{[b]}\right\rangle_\bn=-\left(\frac{\lambda_{\text{so}}}{2}\right)
\frac{\left(Z_\omega^2+3\Delta^2\right)}{(Z_\omega^2+4\Delta^2)Z_\omega\tau}\left\{\hat{\bm{\eta}},\bm{\partial}\hat{\rho}_\veps\right\}\\&+\left(\frac{\lambda_{\text{so}}}{2}\right)\frac{\Delta^2}{(Z_\omega^2+4\Delta^2)Z_\omega\tau}\sum\limits_{a}\hat{\eta}_a\left\{\hat{\eta}_{\overline{a}},\bm{\partial}_{\overline{a}}\hat{\rho}_\veps\right\}\hat{\eta}_{a}.
\end{split}
\en
The remaining step is calculating averages of the type $\left\langle\hat{\eta}_\bp\hat{K}_1\hat{\eta}_\bp\right\rangle_\bn$ and $\left\langle\left[\hat{\eta}_\bp,\hat{K}_1\right]\right\rangle_\bn$ which is straightforward to do with the expression for $\hat{K}_1$ obtained above. Collecting all these terms together in the function ${\cal F}[\hat{K}_1]$ yields the expressions for the coefficients $B$ and $C$ in the main text.

%#############################################################################################################################
%#############################################################################################################################
%#############################################################################################################################
%#############################################################################################################################
%#############################################################################################################################

%$$$$$$$$$$$$$$$$$$$$$$$$$$$$$$$$$$$$$$$$$$$$$$$$$$$$$
\section{Interaction correction to the equation for the density matrix}
%$$$$$$$$$$$$$$$$$$$$$$$$$$$$$$$$$$$$$$$$$$$$$$$$$$$$$

We will use the following auxiliary expressions
\beg\label{Aux1}
\begin{split}
&\left\langle\hat{g}^{(0)}(t,\veps;\bn,\br)\right\rangle_\bn\approx\frac{f_\veps}{2}\hat{\sigma}_0, \\
&\left\langle\hat{g}^{(1)}(t,\veps;\bn,\br)\right\rangle_\bn\approx -eB\frac{\partial f_\veps}{\partial \veps}[\bm{e}_z\times\bm{\mathcal{E}}]\cdot\hat{\bm{\sigma}}
\end{split}
\en
and similarly for the functions
\beg\label{Aux2}
\begin{split}
&\left\langle n_j\hat{g}^{(0)}(t,\veps;\bn,\br)\right\rangle_\bn\approx-\frac{\lambda_{\text{so}}}{2v_F}f_\veps\hat{\eta}_j, \\
&\left\langle n_j\hat{g}^{(1)}(t,\veps;\bn,\br)\right\rangle_\bn\approx-\left(\frac{ev_F\tau}{4}\right){\cal E}_j\frac{\partial f_\veps}{\partial \veps}\hat{\sigma}_0
\\&-\frac{e\tau\lambda_{\text{so}}^2\zeta}{2v_F(1+4\zeta^2)}\frac{\partial f_\veps}{\partial \veps}[\hat{\bm{\sigma}}\times\bm{\mathcal{E}}]_j
\end{split}
\en
Keeping the terms which are linear in electric field, we have
\beg\label{Stel1Approx}
\begin{split}
&\textrm{St}^{[2]}_{\textrm{el}}\left\{\hat{g}^{(1)}\right\}=-\frac{1}{\tau}\int\frac{d\omega}{2\pi}n_i{\cal R}_{ij}^{[1]}(\omega)f_{\veps-\omega}\\&\times
\left\langle n_j\hat{g}^{(1)}(t,\veps;\bn,\br)\right\rangle_\bn=\int\frac{d\omega}{2\pi}n_i{\cal R}_{ij}^{(1)}(\omega)f_{\veps-\omega}\\&\times
\left(\frac{ev_F}{4}{\cal E}_j\hat{\sigma}_0+\frac{e\lambda_{\text{so}}^2\zeta}{2v_F(1+4\zeta^2)}[\hat{\bm{\sigma}}\times\bm{\mathcal{E}}]_j\right)\frac{\partial f_\veps}{\partial\veps}.
\end{split}
\en
Here the $f_\veps$ is determined by the single particle distribution functions in each of Rashba's two bands in equilibrium. 
The second term here is the one which produces the nonzero expectation value of the electron's spin.

Since we are interested in the contribution of the gradient terms only, we will use the expressions above to rewrite the kernel of Eq. (\ref{Stel2}) as follows:
\beg\label{ExpandStel2}
\begin{split}
&\left[\left\langle n_j\hat{g}(t,\veps-\omega;\bn,\br)\right\rangle_\bn\left\langle\hat{g}(t,\veps;\bn,\br)\right\rangle_\bn\right]^{(1)}\\&\approx
-\left(\frac{ev_F\tau}{8}\right)\left(\frac{\partial f_{\veps-\omega}}{\partial \veps}\right)f_\veps{\cal E}_j\hat{\sigma}_0\\&
-\frac{e\lambda_{\text{so}}^2\zeta}{4v_F(1+4\zeta^2)}\left(\frac{\partial f_{\veps-\omega}}{\partial \veps}\right)f_{\veps}[\hat{\bm{\sigma}}\times\bm{\mathcal{E}}]_j\\
&-\frac{e\lambda_{\text{so}}B}{2v_F}f_{\veps-\omega}\left(\frac{\partial f_{\veps}}{\partial\veps}\right){\cal E}_i\hat{\eta}_j\hat{\eta}_i.
\end{split}
\en
Using these expressions along with Eq. (\ref{Stel2}) we have
\beg\label{Stel0a}
\begin{split}
&\textrm{St}^{[1]}_{\textrm{el}}\left\{\hat{g}^{(1)}\right\}=\int\frac{d\omega}{2\pi}n_i{\cal R}_{ij}^{[0]}(\omega)\\&
\left[\frac{\partial f_{\veps-\omega}}{\partial \veps}
\left(\frac{ev_F}{4}{\cal E}_j\hat{\sigma}_0+\frac{e\lambda_{\text{so}}^2\zeta}{2v_F(1+4\zeta^2)}[\hat{\bm{\sigma}}\times\bm{\mathcal{E}}]_j \right)f_\veps\right.\\&\left.+\frac{e\lambda_{\text{so}}B}{v_F}f_{\veps-\omega}\left(\frac{\partial f_{\veps}}{\partial\veps}\right){\cal E}_i\hat{\sigma}_0\right],
\end{split}
\en
where the expression for ${\cal R}_{ij}^{[0]}(\omega)$ is given in the main text. 
Last, there remains one more contribution 
\beg\label{Stel0b}
\textrm{St}^{[2]}_{\textrm{el}}\left\{\hat{g}^{(1)}\right\}=\frac{e\lambda_{\text{so}}B}{v_F}\int\frac{d\omega}{2\pi}n_i{\cal R}_{ij}^{[2]}(\omega)f_{\veps-\omega}\left(\frac{\partial f_{\veps}}{\partial\veps}\right)[\hat{\bm{\sigma}}\times\bm{\mathcal{E}}]_j.
\en

We close this section by looking at Eq. \eqref{g2}.
Since the collision integral is proportional to $\bn$, it is clear that averaging over $\bn$ will render the contributions from the first two terms in ${\cal F}$ to vanish.
Furthermore, only terms $\propto [\bm{\mathcal{E}}\times\hat{\bm{\sigma}}]$ in $\hat{K_2}(\hat{g}^{(1)})$ will contribute, which means that the part of the collision integral proportional to ${\cal R}_{ij}^{[3]}(\omega)$ does not contribute to the spin current.

%#############################################################################################################################
%#############################################################################################################################
%#############################################################################################################################
%#############################################################################################################################
%#############################################################################################################################

%$$$$$$$$$$$$$$$$$$$$$$$$$$$$$$$$$$$$$$$$$$$$$$$$$$$$$$$$$$$$$$$$$$$$$$$$$$$$$$$$$$
\section{Calculation of the collision kernels ${\cal R}_{ij}^{[0]}(\omega)$ and ${\cal R}_{ij}^{[1]}(\omega)$}
%$$$$$$$$$$$$$$$$$$$$$$$$$$$$$$$$$$$$$$$$$$$$$$$$$$$$$$$$$$$$$$$$$$$$$$$$$$$$$$$$$$

In this section our goal will be to compute the angular averages of the diffusion propagators which enter into the expressions for the functions ${\cal R}_{ij}^{[l]}(\omega)$. We start  with the expression for the diffusion propagator
\beg\label{Gamma}
\Gamma(\bn,\bn';\omega,\bq)=\frac{2\pi\delta(\bn-\bn')}{-i\omega+i{\mathbf v}_F\bq+\frac{1}{\tau}}
+\frac{1}{\tau}\frac{\langle\Gamma(\bn,\bn';\omega,\bq)\rangle_\bn}{\left[-i\omega+i{\mathbf v}_F\bq+\frac{1}{\tau}\right]}.
\en
We can now integrate both parts over $\bn$ which yields
\beg\label{avGamma1}
\begin{split}
&\langle\Gamma(\bn,\bn';\omega,\bq)\rangle_{\bn'}=\frac{\tau Y_{q}\left(\tau Y_{q}-1\right)^{-1}}{-i\omega+i{\mathbf v}_F\bq+\frac{1}{\tau}},
\end{split}
\en 
where we use the shorthand notation $Y_q=Y(\omega,\bq)$ and
\beg\label{Yq}
Y_{q}=\sqrt{(v_Fq)^2+\left(-i\omega+\frac{1}{\tau}\right)^2}.
\en
As the next step, averaging Eq. (\ref{avGamma1}) over the directions of the remaining $\bn$ we find
\beg\label{avGamma2}
\begin{split}
&\left\langle \Gamma(\omega,\bq)\right\rangle=\frac{\tau}{\tau Y_{q}-1}.
\end{split}
\en 
Note that from the last expression it follows
\beg\label{RelGam}
\left\langle \Gamma(-\omega,-\bq)\right\rangle=\left\langle \Gamma(\omega,\bq)\right\rangle^*.
\en

\paragraph{Kernel function ${\cal R}_{ij}^{[1]}(\omega)$.} Let us discuss the averages which enter into the expression for ${\cal R}_{ij}^{[1]}(\omega)$, Eq. (\ref{R1}). We have
\beg\label{AvGnj}
\begin{split}
&\left\langle\Gamma\Gamma\right\rangle=\frac{\tau\left(1-i\omega\tau\right)}{Y_q(\tau Y_q-1)^2}, \\
&\left\langle n_x\Gamma\right\rangle=\left\langle\Gamma n_x\right\rangle=\frac{\cos\varphi_\bq}{iv_Fq}\left(1+\frac{i\omega\tau}{\tau Y_q-1}\right), \\
&\left\langle n_y\Gamma\right\rangle=\left\langle\Gamma n_y\right\rangle=\frac{\sin\varphi_\bq}{iv_Fq}\left(1+\frac{i\omega\tau}{\tau Y_q-1}\right),
\end{split}
\en
where we use parametrization $\cos\varphi_\bq=q_x/q$ and $\sin\varphi_\bq=q_y/q$. It becomes clear from these expressions that matrix ${\cal R}_{ij}^{[1]}$ is diagonal since
\beg\label{IntvRij1}
\int\frac{d\varphi_\bq}{2\pi}\left\langle n_i\Gamma\right\rangle\left\langle n_j\Gamma\right\rangle=-\frac{\delta_{ij}}{2v_F^2q^2}\left(1+\frac{i\omega\tau}{\tau Y_q-1}\right)^2.
\en
Using these expressions we obtain
\beg\label{Rij1wom}
\begin{split}
&{\cal R}_{ij}^{[1]}(\omega)=-\left(\frac{\delta_{ij}}{2}\right)\textrm{Im}\int\frac{qdq}{2\pi}\mathcal{D}^R(\omega,\bq)\\&\times
\left\{\frac{Y_q-\frac{1}{\tau}+i\omega}{Y_q\left(Y_q-\frac{1}{\tau}\right)^2}+\frac{(Y_q-\frac{1}{\tau}+i\omega)^2}{v_F^2q^2\left(Y_q-\frac{1}{\tau}\right)^2}\right\}.
\end{split}
\en
In order to compute the momentum integral, we need to find the explicit expression for the retarded part of the bosonic propagator $\mathcal{D}^R(\omega,\bq)$, which satisfies the Dyson equation (\ref{BosonicDyson}). Using the relation between the polarization operators and the fermionic correlators (\ref{PiGRelations}) one finds the following expression for $\mathcal{D}^R(\omega,\bq)$ in terms of the functions considered above:
\beg\label{DR}
\mathcal{D}^R(\omega,\bq)\approx-\nu^{-1}_F\frac{Y_q-\frac{1}{\tau}}{Y_q-\frac{1}{\tau}+i\omega}.
\en
This result has been written in the unitary limit, $\mathcal{D}^R(\omega,\bq)\approx-1/\Pi^R(\omega,\bq)$, which corresponds to taking into account the contributions from the large distances i.e. small momenta [see e.g. Ref. \cite{ANZ2001} for a related discussion on the limits of validity of this expression].
Using these expressions the integral over momentum yields the following result for the function $\mathcal{R}_{ij}^{[1]}(\omega)$:
\beg\label{Res4Rij1}
\begin{split}
&{\cal R}_{ij}^{[1]}(\omega)=\frac{\delta_{ij}}{4E_F}
\left\{\frac{\pi}{2}\left(\frac{6+\omega^2\tau^2}{4+\omega^2\tau^2}\right)\textrm{sgn}(\omega\tau)\right.\\&\left.
+\left(\frac{2+\omega^2\tau^2}{4+\omega^2\tau^2}\right)\arctan(\omega\tau)+\frac{\omega\tau}{4+\omega^2\tau^2}\ln2\right.\\&\left.-\frac{\omega\tau}{4+\omega^2\tau^2}\ln\left(\frac{|\omega\tau|}{\sqrt{1+\omega^2\tau^2}}\right)\right\}.
\end{split}
\en

\paragraph{Kernel function ${\cal R}_{ij}^{[0]}(\omega)$.} We proceed with the calculation of the kernel function ${\cal R}_{ij}^{[0]}(\omega)=\textrm{Im}\overline{\cal R}_{ij}^{[2]}(\omega)$, Eq. (\ref{R2}). In the expressions for the angular averages below we will keep only terms which will give nonzero result upon integration over the directions of the momentum $\bq$:
\beg\label{R2Avs}
\begin{split}
&\left\langle\Gamma n_\alpha\Gamma n_\beta\right\rangle=\frac{i}{v_F}\frac{\partial}{\partial q_\alpha}\left\langle\Gamma n_\beta\right\rangle, \\
&\left\langle n_x \Gamma  n_x\right\rangle=\frac{\sin\varphi_\bq}{Y_q}+\frac{1}{\tau}\left(\frac{\tau Y_q-1}{\tau Y_q}\right)\left\langle n_x\Gamma\right\rangle^2, \\
&\left\langle n_y \Gamma  n_y\right\rangle=\frac{\cos\varphi_\bq}{Y_q}+\frac{1}{\tau}\left(\frac{\tau Y_q-1}{\tau Y_q}\right)\left\langle n_y\Gamma\right\rangle^2.
\end{split}
\en
Collecting all these terms, we find
\beg\label{Rij2wom}
\begin{split}
&{\cal R}_{ij}^{[0]}(\omega)=\left(\frac{\delta_{ij}}{2}\right)\textrm{Im}\int\frac{qdq}{2\pi}\mathcal{D}^R(\omega,\bq)\\&\times
\left\{\frac{Y_q-\frac{1}{\tau}+i\omega}{Y_q\left(Y_q-\frac{1}{\tau}\right)^2}+\frac{(Y_q-\frac{1}{\tau}+i\omega)^2}{v_F^2q^2\left(Y_q-\frac{1}{\tau}\right)^2}\right\}.
\end{split}
\en
From this result we conclude that 
\beg\label{R12ijrel}
{\cal R}_{ij}^{[0]}(\omega)=-{\cal R}_{ij}^{[1]}(\omega)\equiv\delta_{ij}{\cal R}(\omega).
\en
Hence, we use this expression along with Eq. (\ref{Res4Rij1}) to write Eq. (\ref{FinalJik}). 

Finally, we turn our attention to the expression for ${\cal R}_{ij}^{[2]}(\omega)$, Eq. (\ref{R0R2Again}). We find
\beg\label{R2ReRij2}
\begin{split}
&{\cal R}_{ij}^{(2)}(\omega)=\frac{\delta_{ij}}{4E_F}
\left\{\left(\frac{6+\omega^2\tau^2}{4+\omega^2\tau^2}\right)\ln\left(\frac{v_Fq^*}{|\omega|}\right)
\right.\\&\left.+\left(\frac{2+\omega^2\tau^2}{4+\omega^2\tau^2}\right)\ln\left(\frac{v_Fq^*\tau}{2\sqrt{1+\omega^2\tau^2}}\right)\right.\\&\left.
-\frac{\pi|\omega\tau|}{2(4+\omega^2\tau^2)}+\frac{\omega\tau}{4+\omega^2\tau^2}\arctan(\omega\tau)\right\}.
\end{split}
\en
Here we introduced the momentum cutoff $q^*\approx p_F$.
\end{appendix}

\bibliography{biblio}

\end{document}